\documentclass[11pt]{article}
\usepackage{graphicx}
\usepackage{savesym}
\usepackage{amsmath}
\savesymbol{iint}
\usepackage{txfonts}
\restoresymbol{TXF}{iint}

\usepackage{amssymb,amsfonts,verbatim}
\usepackage{url}
\usepackage{latexsym}
\usepackage{natbib}
\usepackage{aas_macros}

\def\x{{\mathbf x}}
\def\k{{\mathbf k}}

\def\E{{\hbox{I\kern-.2em\hbox{E}}}}  
\def\R{{\mathbb{R}}}

\newcommand{\refeq}[1]{~(\ref{#1})}            
\def\ds#1{\displaystyle{#1}}

\def\1{\ifmmode{\rm {I}\mkern-10.1mu
{1}\mkern0.5mu}\else{\rm {I}\kern-.56em
{1}\hskip0.5pt\ }\fi\relax}

\begin{document}
\title{Quantifying and containing the curse of high resolution coronal imaging}

\author{V. Delouille$^{1,3}$ \and P.Chainais$^2$ \and J.-F. Hochedez$^{2}$}
\footnotetext[1]{Royal Observatory of Belgium, Circular Avenue 3, B-1180 Brussels, Belgium}
\footnotetext[2]{LIMOS, Universit\'e Blaise Pascal, Clermont-Ferrand, France}
\footnotetext[3]{Corresponding author. Email: v.delouille@sidc.be}

\date{30 July 2008}

\maketitle

\begin{abstract}

Future missions such as Solar Orbiter (SO), InterHelioprobe, or Solar Probe aim at approaching the Sun closer than ever before, with on board some high resolution imagers (HRI) having  a subsecond cadence and a pixel area of about $(80km)^2$ at the Sun during perihelion.  In order to guarantee their scientific success, it is necessary to evaluate if  the photon counts available at these resolution and cadence will provide a sufficient signal-to-noise ratio (SNR). 

For example, if the inhomogeneities in the Quiet Sun emission prevail at higher resolution, one may hope to locally have more photon counts than in the case of a uniform  source. 
It is relevant to quantify how inhomogeneous the quiet corona will be for a pixel pitch that is about 20 times smaller than in the case of SoHO/EIT, and 5 times smaller than TRACE.

We perform a first step in this direction by analyzing and characterizing the
spatial intermittency of Quiet Sun images thanks to a multifractal analysis. 
 We identify the parameters that specify the scale-invariance behavior. This identification allows next to select a family of multifractal processes, namely the Compound Poisson Cascades, that can synthesize artificial images having some of the scale-invariance properties observed on the recorded images. 

The prevalence of self-similarity in Quiet Sun coronal images makes it relevant to study the ratio between the SNR present at SoHO/EIT images and in coarsened images. SoHO/EIT images thus play the role of \lq high resolution' images, whereas the \lq low-resolution' coarsened images are rebinned so as to simulate a smaller angular resolution and/or a larger distance to the Sun.
For a fixed difference in angular resolution and in Spacecraft-Sun distance,
we determine the proportion of pixels having a SNR preserved at high resolution  given a particular increase in effective area. 
If scale-invariance continues to prevail at smaller scales, the conclusion reached with SoHO/EIT images can be transposed to the situation where the resolution is increased from SoHO/EIT to SO/HRI resolution at perihelion.
\end{abstract}

\vspace{0.5cm}
\vspace{0.5cm}

\section{Introduction}

Many fine-scale structures in the corona do not seem to be well resolved by current imaging telescopes. Filamentary and threaded patterns are observed in coronal loops~(e.g.~\citet{2007ApJ...661..532D}); in the Quiet Sun (QS), small dynamical events such as brightenings or blinkers point at unresolved sub-structures.
Having sufficient resolution is also necessary to assess
 whether nanoflares occurring in QS may explain coronal heating
~\citep{1988ApJ...330..474P,1998ApJ...501L.213K,2007ApJ...659.1673A,2001A&A...373..318M,2002ESASP.506..501B}.
Several studies~\citep{1998A&A...336.1039B,1998ApJ...501L.213K,2000ApJ...544..550A} present evidence in favor of a turbulent mechanism, with individual dissipative structure far below the instrumental resolution limit.

Scale-invariance has been observed down to the smallest reachable scales, namely down to the PSF dimensions of today's highest resolution instruments. This paper studies this scale-invariance (or self-similary) property in the case of a SOHO/EIT data set, first by computing the Fourier spectrum, and next by performing a multifractal analysis.
But, should this property persist to even smaller scales? To answer this question, we can only rely on current observation extrapolations and on physical theory and modeling. We now briefly discuss some of those approaches. 

A first element stems from the consistency between quantities observed by both EIT and TRACE, albeit at resolutions differing by a factor five in 1D
($\times 25$ in 2D). For example, \citet{2002ApJ...568..413B} show that EIT and TRACE studies agree about the distribution of energy released by flare-like events in the Quiet Sun. Power-law distributions fit both data, and they find similar values for their slopes. 
Along a similar line, a H\"older analysis in~\citet{dch:spatial_noise} evidences a continuity between supra- and intra-pixel scales. Intra-pixel variability is therein estimated using a one-minute cadence EIT data set, for which solar rotation induces a displacement (from one image to the next) that is less than the pixel.
To go to smaller scales, note first that regions of strong magnetic field, such as bright points, exhibit brighter EUV emission, while regions of weak magnetic field lead to a darker emission. Hence scale-invariance in the EUV emission might follow from similar properties in the magnetic field strength. 
The tectonics model of~\citet{2002ApJ...576..533P} suggests a highly fragmented photospheric magnetic configuration in the quiet network, where the fundamental units of flux  are intense flux tubes having a 100km diameter. 
Finally, much literature has been devoted to the study of distribution of flare properties, see e.g.~\citep{1998A&A...336.1039B,2000ApJ...535.1047A,2004ApJ...603L..61V}. Power-law distributions for e.g.~the energy release and volume occupied by flares might be explained by self-organization and the cascading nature of flare activities. Such statistical theory of flare activity are physically motivated by the turbulent nature of the solar corona~\citep{2003matu.book.....B,1996ApJ...457L.113E} where the dissipative scale is estimated to be of a few meters~\citep{1994SSRv...68...97E}.

One notices that the above reasoning relies on rather hypothetical guesses, and this is an additional motivation to actually plan high resolution observations of the QS in coronal lines. We are also aware that the observed intensity in a given EUV optically thin spectral line entangles the temperature and the emission measure. The above issue is worsened by the possible contamination from other lines in the passband, and by the projection effects. It remains nevertheless of interest to prepare the conceptual tools that will enable the analysis of better quality data when they come.


Beyond the Fourier spectrum,  a multifractal analysis allows to further characterize the spatial scale-invariance, or more generally the so-called \lq intermittency' of a stochastic process~\citep{cd95,frischbook95,amr97,chap99}. It has been extensively used for the statistical modeling of turbulent flows~\citep{frischbook95}, Internet traffic~\citep{fgw98}, natural~\citep{tmp98,chainaistpami07} and meteorological images~\citep{rad00,gtyh07}, as well as ionospheric indices~\citep{PhysRevLett.76.4082}.
 
 In solar physics, studies on the fractal dimension was achieved for Quiet Sun EUV network~\citep{1998A&A...335..733G}, for the spatial extend of nanoflares events~\citep{2002ApJ...572.1048A}, and for active regions~\citep{2005ApJ...631..628M}. 
  \citet{law-cad-ruz:multiplic} present a  multifractal analysis of photoelectric images of line-of-sight magnetic fields in solar active regions and quiet photosphere. They consider a multiplicative cascade approach in order to infer a 
 scale-invariant rule for the Ohmic dissipation measure. This rule is then used to re-estimate with greater accuracy  the multifractal spectrum.
 Multifractal analysis has been also applied to the analysis of solar magnetograms~\citep{2005SoPh..228....5G,2005SoPh..228...29A}, and of the temporal variation of the emission observed in several RHESSI X-ray energy bands~\citep{2007ApJ...662..691M}.

In this work, we aim at generating synthetic images at EIT resolution that capture several statistical properties of Quiet Sun images. To do this with minimum a priori, we propose to exploit the statistical scale invariance observed in data. We first perform a multifractal analysis. It yields a set of parameters, namely the multiscaling exponents, that quantitatively describe this scale invariance. Next we aim at finding a family of stochastic processes that obey the same statistical property with the same set of parameters, thus injecting minimum a priori in the model. The corresponding synthetic images are exempt of spurious artifacts such as  cosmic ray hits, projection effect, or error sources in the data.
This is the first achievement necessary in the elaboration of a way to create synthetic images at arbitrary resolution. The analysis and parameter identification step is essential in order to know how to extrapolate the properties of EIT images at higher resolution. Moreover, our methodology for synthesizing Quiet Sun EUV images is new: having observed certain properties on EIT images, the associated parameters are then used within a multifractal stochastic process that can simulate EIT images.

The are two main applications of this procedure in solar physics.
First, artificial EUV images can be used for the testing and calibration of automatic feature finder procedure, see e.g.~\citet{2007A&A...464.1107G}. 
Second, and most importantly for the purpose of this paper, since QS images exhibit self-similarity behavior that can be reproduced with a multifractal stochastic process, it makes sense to study the relationships between intensities across different scales of observations. More precisely, we can look at the statistics of the ratio between intensity values at full resolution and at a rebinned version of the original images. The conclusion reached here can be transposed to higher resolutions as long as the scale-invariance continues to prevail. Our study therefore provides some guidance for radiometric studies related to future high resolution missions. The latter are much needed since the spatial resolution of current telescopes precludes definite conclusion about the fundamental processes that determine the existence and the underlying physics of the transition region and of the  quiet corona. In the following, we are considering in more details the case of High Resolution Imagers (HRI) on board Solar Orbiter (SO).

HRIs on board SO will produce at perihelion images having a pixel area at the Sun of $(80\mbox{km})^2$.
 Evaluating the loss in SNR when going from low to high resolution is necessary in order to guarantee the scientific success of this high resolution mission.
Indeed, the smaller the area covered by a pixel, the lesser the signal, and the fainter and more dynamical the target. 
Similarly, the higher the cadence, the shorter the exposure time, and the lesser the signal again.

Considering that the energy $E$ emitted from the Sun follows a Poisson process, and hence that 
$\mbox{SNR} \leq E/\sqrt{E}$, $\mbox{SNR}^2 \leq E$, 
\cite{2001ESASP.493..245H} describe the relationship between the quality of  observations (left hand side) and the experimental conditions (right hand side) as follows:
\begin{equation}
\label{E:SNRineq}
\frac{\text{SNR}^2}{A_{\mbox{sun}}  T_{\mbox{ExpTime}}} \leq \frac{A_{eff} L}{d^2}~,
\end{equation}
where $A_{sun}$ is the Sun area covered by one pixel, $T_{ExpTime}$ is the exposure time, $A_{eff}$ is the effective area of the telescope (in $m^2$),  $L$ is the radiance (in $W sr^{-1} m^{-2}$), and $d$ is the distance between the spacecraft and the Sun.

When comparing the situation of EIT with the one foreseen for an HRI on board SO, we observe that getting closer to the Sun provides a factor $d^2 = 25$, but since $A_{sun}$ is divided at the same time   from $(1800km)^2$ (EIT) to $(80km)^2$ (HRI), 
the effective area $A_{eff}$ needs to be enhanced by a factor of $(1800/80)^2 /25 \simeq 20$ or more  in order to preserve a SNR for a uniform source in space. Moreover, if one takes into account the temporal variability of solar features, this factor gets even larger: ~\citet{1998A&A...336.1039B} showed that the typical duration of an event was proportional to its size. If this argument still holds at a scale of $(80\mbox{km})^2$, HRI should have an exposure time that is $(1800/80)^2$ times less than in the case of EIT. The effective area should then increase by a factor 
$\left(\frac{1800}{80}\right)^4/25 \simeq 10^4$
in order to have a SNR similar to EIT images.

Our main  contribution in this paper is to refine this argument by taking into account the inhomogeneities present in the solar corona. Indeed, these might provide an independent help in the quest for a sufficient SNR, with subsets of high-resolution pixels likely to carry a large part of the energy present at a lower resolution.

Towards quantifying the needed gain in $A_{eff}$, we first re-write~(\ref{E:SNRineq}) in a more general form as
 \begin{equation}
 \label{E:SNRineq2}
 SNR^2 \leq \frac{A_{eff}}{d^2}  \int_{T_{ExpTime}}\int_{A_{sun}}  L(s,t)dtds~,
 \end{equation}
which now considers the time dynamics and possible spatial inhomogeneities in the emission process. 
In this paper, we consider a fixed exposure time and a fixed effective area, and 
we investigate  how the spatial distribution of the radiance will affect the SNR in the case of Quiet Sun data.
As mentioned above, the temporal variability is also necessarily non-uniform;  the methods presented here will be adapted in the future to
see how this variability affects positively the SNR in case of a high cadence instrument. 

We  are interested in the evolution of the quantity 
\begin{equation}
\label{E:eqQ}
Q=  \frac{\int_{A_{sun}} L(s)ds}{d^2}~,
\end{equation}
with a higher resolution. 
More precisely, we simulate the degradation in resolution starting from EIT images using an appropriate coarsening of the data.
Taking  the ratio between the value of $Q$ obtained at low (coarsened) and high (EIT original)  resolution, we obtain an estimate of the gain in effective area needed in order to keep at high resolution the SNR (as defined in~\refeq{E:SNRineq}) larger or equal to its value at low resolution. 
 
 If the scale-invariance continues to prevail down to a $(80km)$ scale, our conclusion about the needed gain in $A_{eff}$ may be transposed to the situation where the resolution is increased from EIT to HRI resolution at perihelion.

In future work, we aim at generating images directly at HRI resolution at perihelion. However, additional difficulties arise in this case because not only must the multifractal spectrum  be respected, but the histogram of rebinned images must fit the histogram of EIT images as well. 
 Indeed, the multifractal stochastic model used in this paper to generate QS images does not constrain the simulated images to have the same histogram as the real ones. (Both histograms are however fairly close for images generated at the same scale of observation as the physical ones.)

This paper is structured as follows.
Section~\ref{S:obs} describes the observations considered here with a general overview of noise sources present in EIT images, followed by a presentation of the data set used in this paper. 
In Section~\ref{S:waveletMF} we first recall the  basics behind wavelet-based multifractal analysis. Next we present our multifractal analysis of Quiet Sun images and discuss its limitation.
 With the set of parameters identified in Section~\ref{S:waveletMF}, we are able to synthesize in Section~\ref{S:synthe} Quiet Sun-like images. 
The real and artificial EIT data sets
 are used in Section~\ref{S:fromEIT} to study  the effects of the transition from low to high resolution when taking into account the multifractal behavior of the observations.
 We then discuss the implications of our study in the construction of high-resolution instruments.

\section{Observations}
\label{S:obs}

 We first recall the main sources of errors present in EIT images.
After having described our data set, we discuss the shape of its histogram and power spectrum.

\subsection{Noise in EIT images}

Several sources of errors are contaminating the recording of incident electromagnetic flux in EUV images.
In a simplified description of the process converting electromagnetic flux into digital numbers, photons first impinge the optical system. The photon interaction process can be described by a Poisson distribution, and is most of the time the dominant source of noise.  Second the Point Spread Function (PSF) acts as a blurring operator, and introduces spatial average of unresolved features. This is at the origin of the so-called \lq spatial noise' discussed in~\citep{dch:spatial_noise}.
Third, a spectral selection is operated on the signal before it reaches the CCD detector. 
The latter presents inhomogeneities in its response from one pixel to the next; these have to be  flat-field corrected, usually using the solar software (ssw) library. Fourth, when
 electronics next converts photon counts to digital numbers some read-out noise is introduced.
Finally, note that SoHO/EIT records the image in a lossless way, whereas most of the STEREO/SECCHI/EUVI images are recorded using a lossy compression scheme. Such scheme introduces artifacts, especially in regions with low photon counts.
For a more detailed description of these types of noise, we  refer to~\citet{dch:spatial_noise}.

In the following, we consider that the Poisson noise is dominant, and therefore that $\mbox{SNR}^2 \leq E,$ where $E$ is the energy emitted from the Sun. In Section~\ref{S:fromEIT}, we will talk indifferently about \lq intensities' and \lq SNR'.

\subsection{Data set description}
\label{S:dataset}

Our study is based on full-Sun images taken between 2 January and 28 December  1997 by EIT~\citep{1995SoPh..162..291D} onboard SoHO in the Fe XII (19.5 nm) line, which forms at 
$1.6\times 10^6$K. We selected a first set of $ 1024\times 1024$ full Sun images
recorded with at least 5 days parsing, so as to ensure statistical independence between the images.

\begin{figure}
\begin{center}
\includegraphics[height=70mm]{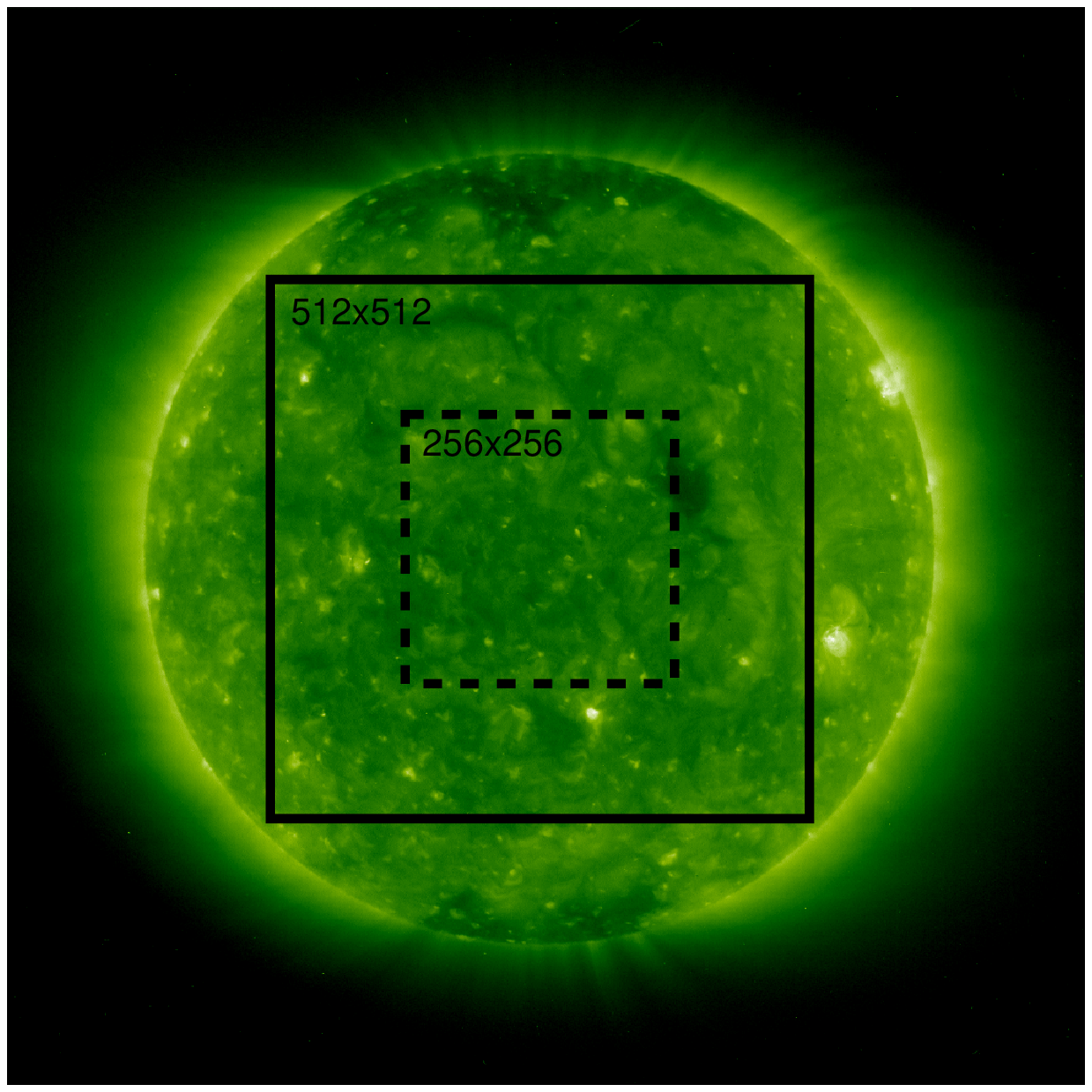}
\end{center}
\caption{\label{F:exempleQS97zone} 
Example of a Quiet Sun images seen by EIT in 19.5~nm, with indication of the squared area taken into account. The multifractal analysis characterizes the region within the square of size $256\times 256$.}
\end{figure}

These images contain no missing blocks, and are processed with the \verb+eit_prep+ procedure of the solar software (ssw) library.
Further, in order to limit the shortening effect at the limb to roughly $15\%$, we do not consider the full Sun images but rather disks $\Omega$ of a half solar radius wide centered on the Sun, see Figure~\ref{F:exempleQS97zone}.  

Next, we want to exclude disks containing active regions.  To do so, 
we keep only disks with intensities $I$ such that the probability $P(I >U)\leq 10^{-4}$ for a given threshold $U$, defined as the quantile such that $P(I>U) = 10\%$ for an image containing one active region.
  We obtain a set of 54 disks centered on the Sun, all having the same field of view. 
Note that we did not attempt to erase cosmic rays, and hence no interpolation or denoising method has been used.

  To perform the wavelet-based multifractal analysis in Section~\ref{S:waveletMF} below, we  consider
  collectively the 54 squared images of size $512 \times 512$ containing the solar disks of 1/2 radius wide. In fact, in order to deal with border effect in the wavelet analysis, we consider only the wavelet coefficients associated to the squared image of size $256\times 256$. In other words, the multifractal analysis is performed on the $512 \times 512$ images, but we keep only results exempt of any border effect, hence in the end, it is the squared image of size $256\times 256$ lying within the disk of a half radius wide which is characterized by our analysis, see Figure~\ref{F:exempleQS97zone}.

\begin{description}
\item[{\bf Remark}] We considered EIT rather than TRACE images for the following reasons. EIT images have an always constant exposure time, and the EIT archive has a regular sampling in time and space. It allows to collect images having the same Field of View, and to limit the shortening effect at the limb. Moreover, the compression algorithm of the TRACE images would introduce another artifacts in the analysis. 
\end{description}


\subsection{Histogram}
\label{histoI}

The solar disks data set described above contains a sample of $5.8 \times 10^6$ pixels. The resulting histogram on Figure~\ref{F:histoIlogI97_powerspect}(a) 
shows a highly non-Gaussian distribution, with the logarithmic representation exhibiting a straight line for large intensities indicative of a power-law behavior. The corresponding power-law index is estimated to 4.9, which is in good agreement with \cite{2000ApJ...544..550A}.

\subsection{Spatial power spectrum}
\label{S:power_spectrum}
\begin{figure*}
\begin{center}
\begin{tabular}{ccc}
\includegraphics[height=55mm]{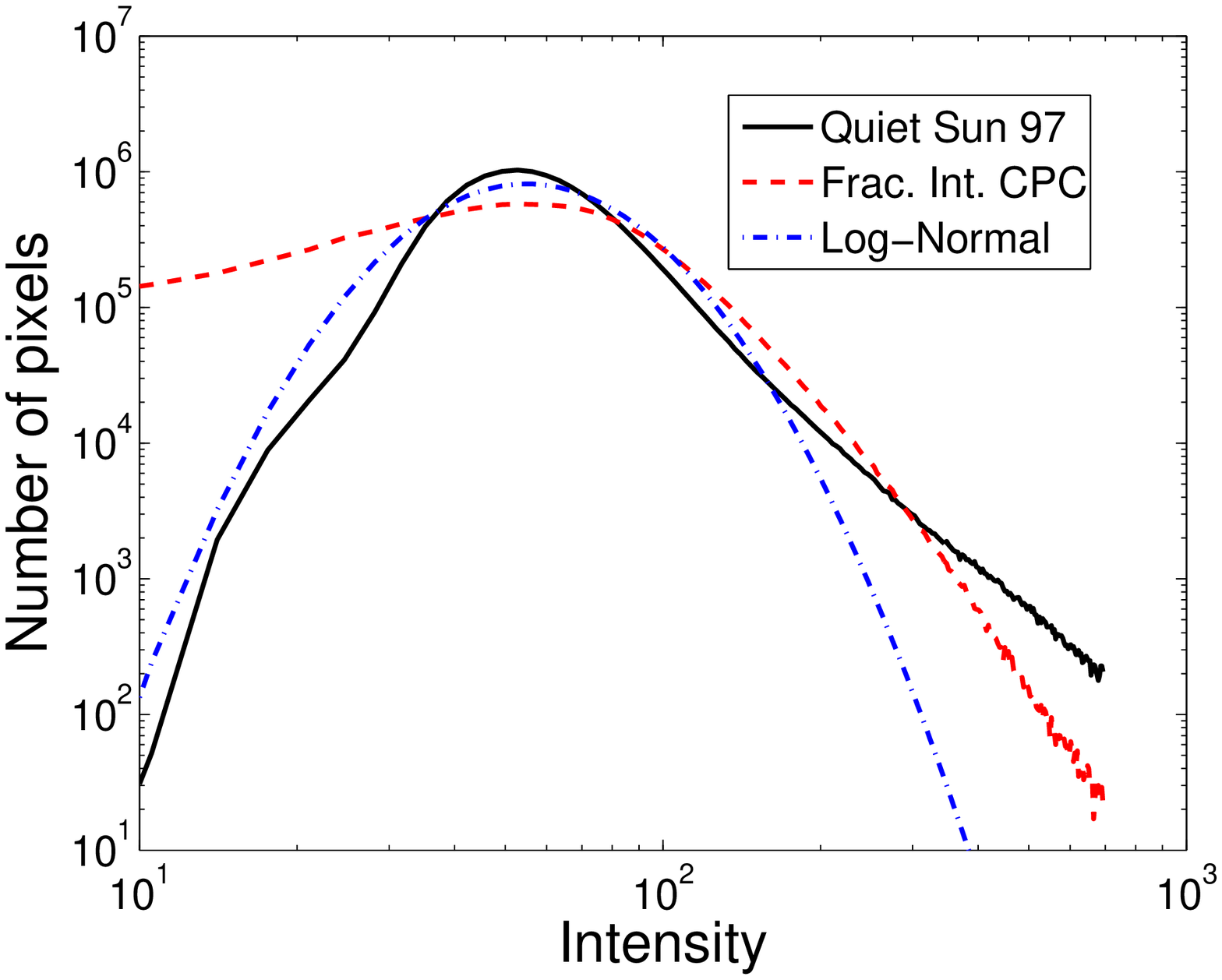}
&
\includegraphics[height=55mm]{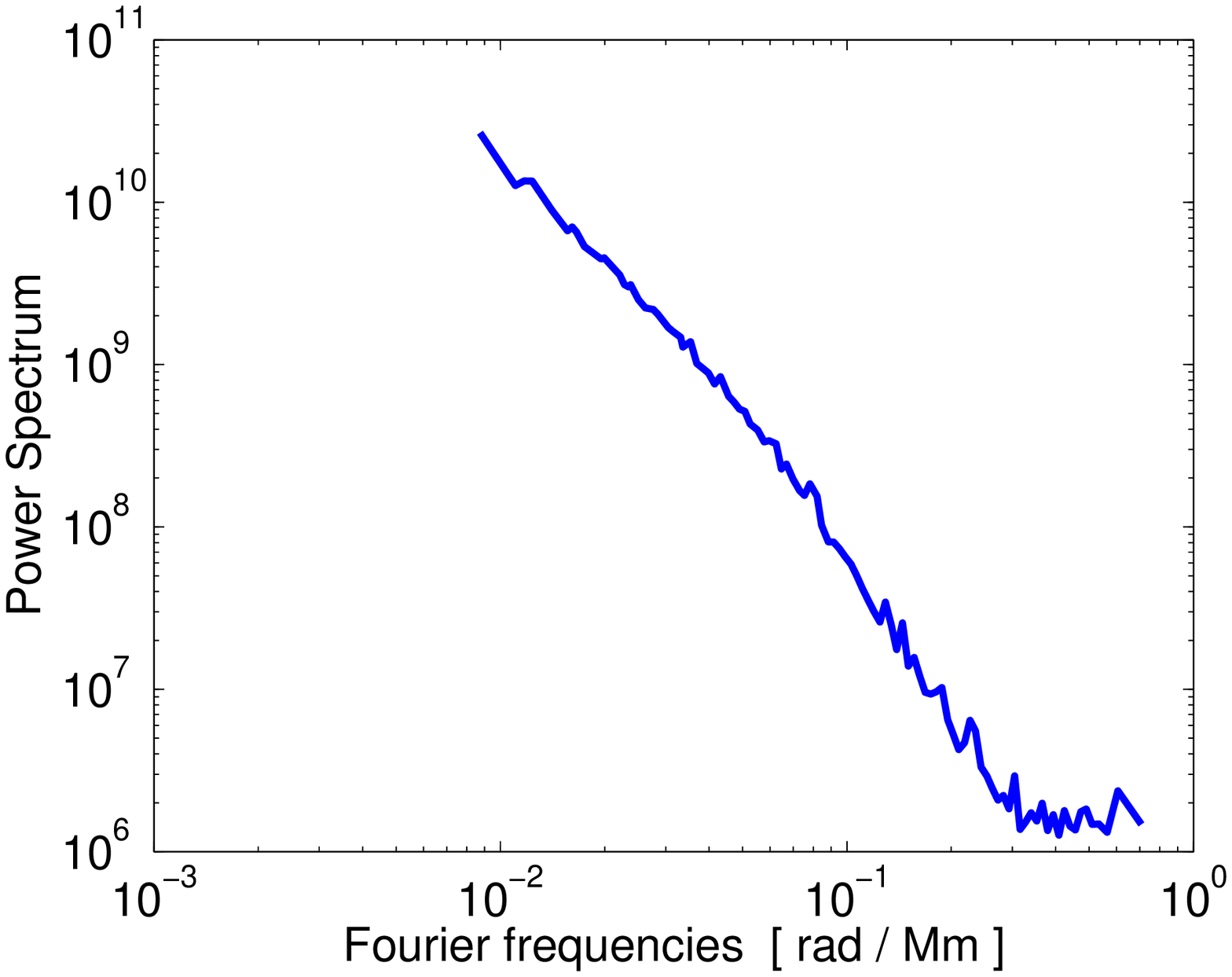}\\
(a) & (b)&(c)
\end{tabular}
\end{center}
\caption{\label{F:histoIlogI97_powerspect} 
(a) Histogram of original data set in log-log scale, together with a lognormal distribution having the same mean and variance, and the histogram of a set of 54 artificial images generated by a multiplicative cascade model. 
(b)Omnidirectional  power spectrum. For Fourier wavenumbers  $0.03<||k||< 1.12$ rad/Mm, the spectrum exhibits a power law with index equal to -2.8 (estimation using least squares)}
\end{figure*}

Fourier spectral density of  Quiet Sun images gives indication about the scale invariance nature of the process.
Individual power spectra are computed for each of the 54 images of size $256 \times 256$. We check the validity of the time-invariance and isotropic assumption and next we compute the 
omnidirectional power spectrum by averaging over all wavenumbers $\k=(k_1,k_2)$ having a same norm $||\k||$, see Figure~\ref{F:histoIlogI97_powerspect}(b). There,  the Fourier wavenumbers $\k$ are expressed in rad/Mm, where we consider that  one pixel corresponds to a scale size of $1.8\mbox{Mm}$ on the Sun.

The  azimuthally integrated spectrum  decreases for Fourier wavenumbers $\k< 1.1~\mbox{rad/Mm}$, and becomes constant thereafter.
 This upper bound corresponds to scale sizes of 5700 km 
and is close to the Nyquist number equal to $2*G$, where $G$ is the
effective resolution of the instrument ($G=1.8~\mbox{Mm}$ for EIT).

For almost two orders of magnitudes (between $0.03 $ and $1.12~\mbox{rad/Mm}$), the spectrum follows a power law with index equal to  $-2.8$.
This value is close to the $-2.7$ index value found by~\citet{1997A&A...320..993B} for Yokhoh/SXT images showing quiet disc corona, and is similar to the index value of $-3$ reported by \citet{1993ApJ...405..767G} for active regions observed by the Normal Incidence X-ray telescope~\citep{1990Natur.344..842G}. On a three hours data set of EIT $19.5\mbox{nm}$ images,~\citet{1998A&A...336.1039B} found a lower value of $-2.52$.

The power law density exhibited by the omnidirectional power spectrum is an indicator of the scale invariance, or self-similarity property of Quiet Sun images in a certain range scale.
For high wavenumbers the Fourier spectrum represents mostly white noise, hence the 
flattening for half a decade observed between $0.3~\mbox{rad/Mm}\leq k \leq 0.7~\mbox{rad/Mm}$ in~Figure~\ref{F:histoIlogI97_powerspect}.

The next section investigates more deeply this  self-similarity property through a
 wavelet-based multifractal analysis.

\section{Wavelet based multifractal analysis}
\label{S:waveletMF}

\subsection{Wavelet analysis}
We know from previous section that the (second
order) power spectrum obeys a power law behavior.  Moreover, the scale invariance property is usually connected to some spatially intermittent behavior. Thus, a refined
analysis of the scale invariant behavior of Quiet Sun images requires
the study of higher order quantities  based on quantities that are spatially localized, which is not the case of Fourier modes. We use here the discrete wavelet transform  (DWT) coefficients derived from an orthonormal wavelet basis~\citep{Daub10,mallat:tour}.

We consider the 54 EIT images data set described in Section~\ref{S:dataset}.
Each of these images is decomposed using the tensor product algorithm, which applies  one-dimensional (1D) DWT on the column of an image, followed by another 1D transform on the rows of the resulting coefficient matrices. We thereby obtain 3 types of detail coefficients, that  give respectively information about vertical, horizontal and diagonal variations in the image.
The 1D filter is  the minimal phase  Daubechies filter with 2 vanishing moments. It was computed with the \emph{Rice Wavelet toolbox} \\
(\verb+http://www-dsp.rice.edu/software/rwt.shtml+).

\begin{figure}
\begin{center}
\includegraphics[height=8.3cm]{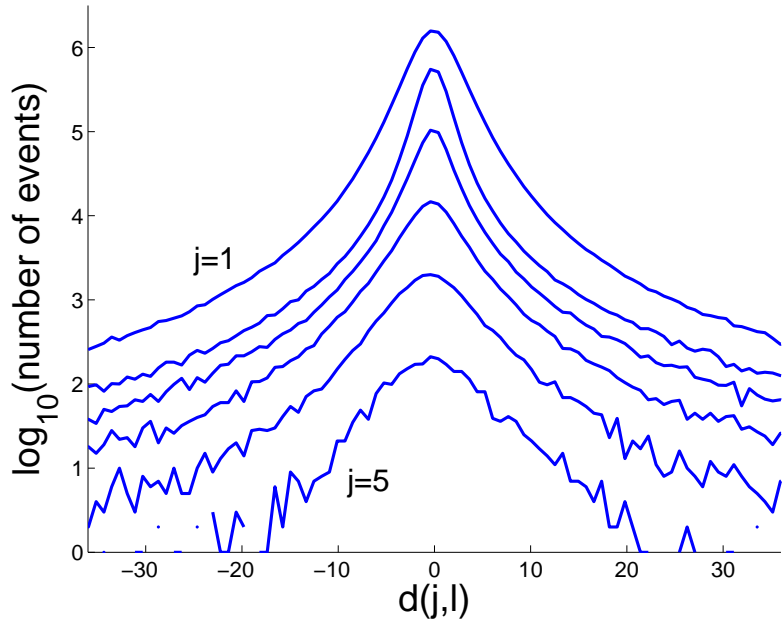}
\end{center}
\caption{\label{F:histoDWTj}  Histograms of 2D discrete wavelet coefficients computed from 54 Quiet Sun images taken by EIT at 19.5~nm during the year 1997.}
\end{figure}

Figure~\ref{F:histoDWTj} shows the histograms of the corresponding 2D wavelet coefficients at different scales. Note that the horizontal, vertical, diagonal coefficients are considered all together. There, the probability density functions of wavelet coefficients evolve from nearly Gaussian at larger scales $a$ ($a=2^j=32,\, j=5 $) to far from Gaussian at smaller scales ($a=1$).
 This observation combined to the power law spectrum illustrates the spatial \emph{intermittency}, or patchiness,  of the Quiet Sun EUV images.
Similarly, ~\cite{2000ApJ...544..550A} noticed that at smaller scales one observes stronger departures from Gaussian statistics.

\subsection{Principle of multifractal analysis}
\label{S:principleMF}

The purpose of multifractal analysis is twofold. First, when dealing with functions or realizations of stochastic processes, it can be used to characterize the relative importance of singularities in the data.  This is quantitatively done by identifying the so-called multifractal (or singularity) spectrum $D(h)$ which can be roughly interpreted as a distribution of singularities characterized by the so-called H\"older exponent $h$, see Appendix~\ref{S:legendre}. The closer $h$ to zero, the stronger the singularity. Second, when dealing with scale invariant stochastic processes in particular, the multifractal analysis can accurately characterize the statistical structure of the process. Indeed, it appears that the singular behavior of realizations (described by $D(h)$) of a self-similar process can generally be connected to the scaling behavior of the so-called structure functions $S(q,a)$. A possible definition of these structure functions is based on the use of a DWT. Then  $S_{DWT}(q,j)$ is computed as the empirical moment of order $q$  of the wavelet coefficient modulus~\citep{abry:revisiting,jaf97ab,riedi03}:
\begin{equation}
  \label{strucfun}
  S_{DWT}(q,j)=\frac{1}{n_j}\sum_{l=1}^{n_j} |d(j,l)|^q\sim 2^{j\zeta(q)}
\end{equation}
where $n_j$ is the number of wavelet coefficients $d(j,l)$ at scale $a=2^j$. For a multifractal process, the $S_{DWT}(q,j)$ obey a powerlaw scaling behavior $S_{DWT}(q,j)\sim a^{\zeta(q)}\sim 2^{j\zeta(q)}$, at least for some range of scales $a$ and orders $q$'s. For a large number of self-similar processes, it can be established that the multiscaling exponents $\zeta(q)$ and the multifractal spectrum $D(h)$ are linked through a Legendre transform, see Appendix~\ref{S:legendre}. The function $\zeta(q)$ is a signature of the scale invariance property of the process under study. 
For a stochastic multifractal process, they contain information on the way the distributions of wavelet coefficients change from larger to smaller scales. This is the reason why we rather talk of `multiscaling'~\citep{cd95,chainaistpami07}. The set of exponents $\zeta(q)$ is used as a set of parameters in the model described in Section~\ref{S:synthe}.  In our application, $S_{DWT}(q,j)$ is obtained by summing over the wavelet coefficients of the 54 images that composes our data set. This method is fast and efficient for positive values of $q$ but becomes numerically unstable for $q<0$ since the most probable value of the $d(j,l)$ is zero. Although in principle it is better to know $\zeta(q)$ for all values of $q$, in our case values of $q<0$ do not bring fundamental information. 
The knowledge of $\zeta(q)$ for $q\geq 0$ is sufficiently informative to constrain our model in Section~\ref{S:synthe}.

\subsection{Multifractal analysis of Quiet Sun images}
\label{QSMFA}

\begin{figure*}
\begin{center}
\begin{tabular}{ccc}
	\includegraphics[height=40mm]{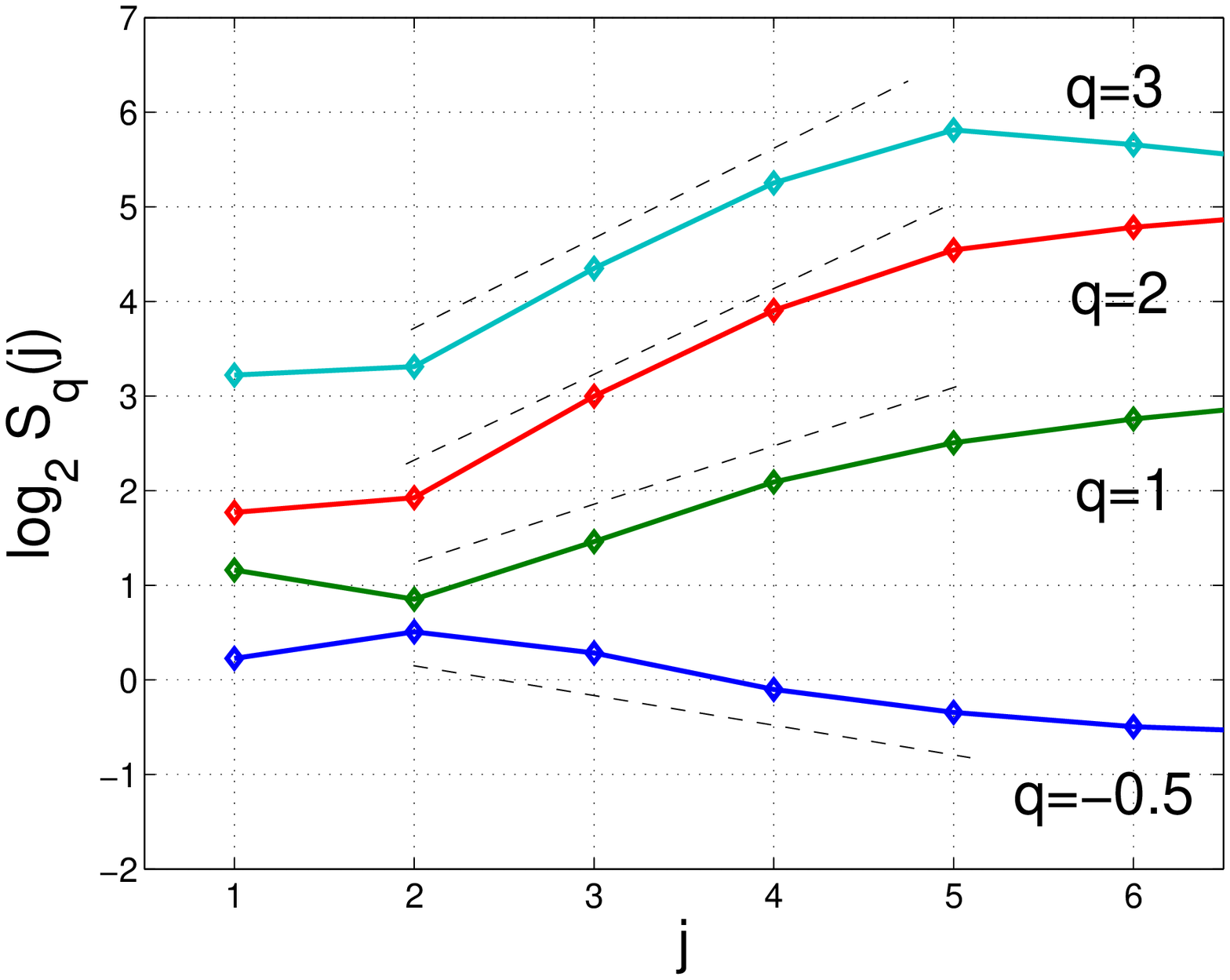}
	&
	\includegraphics[height=40mm]{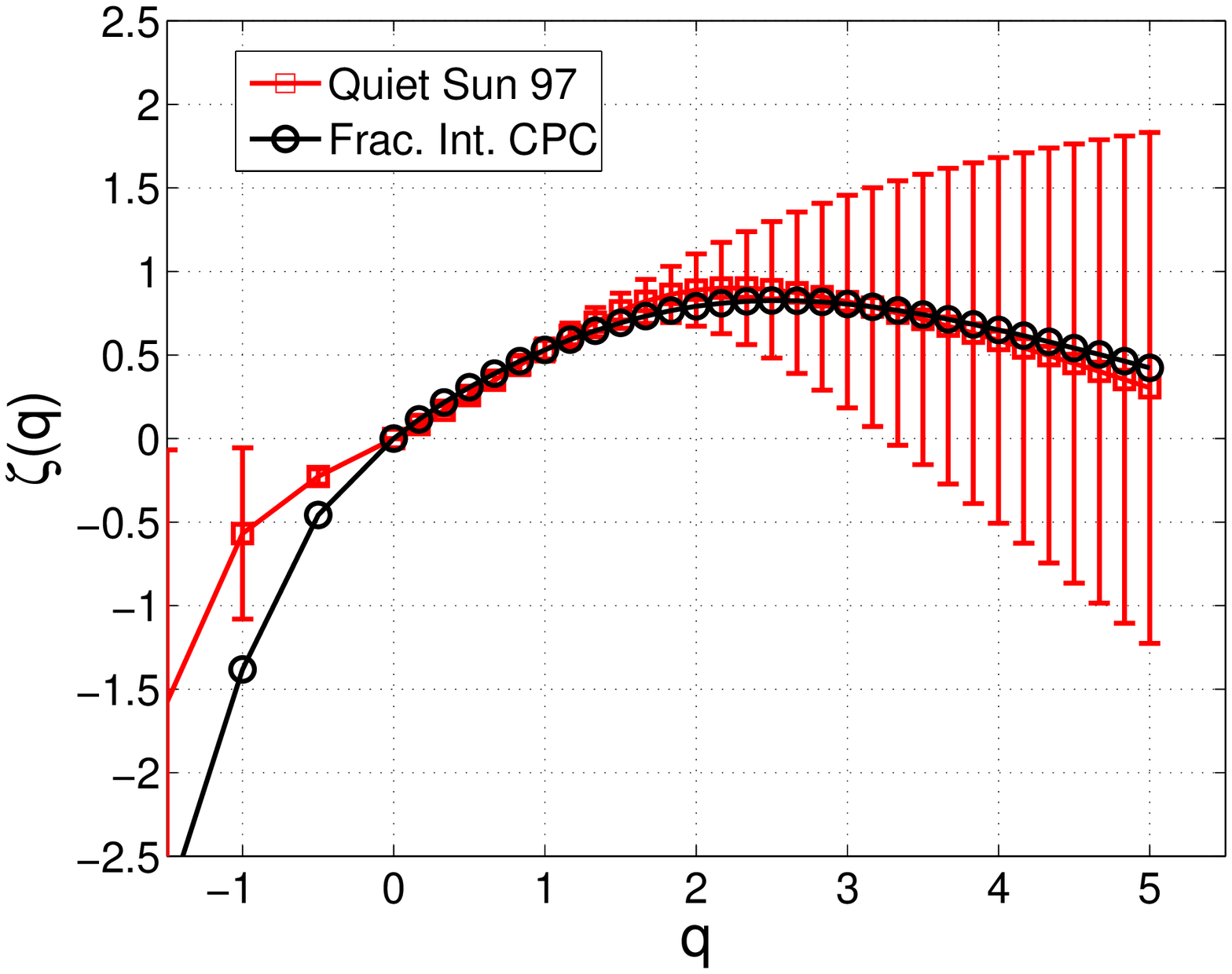}
	&
	\includegraphics[height=40mm]{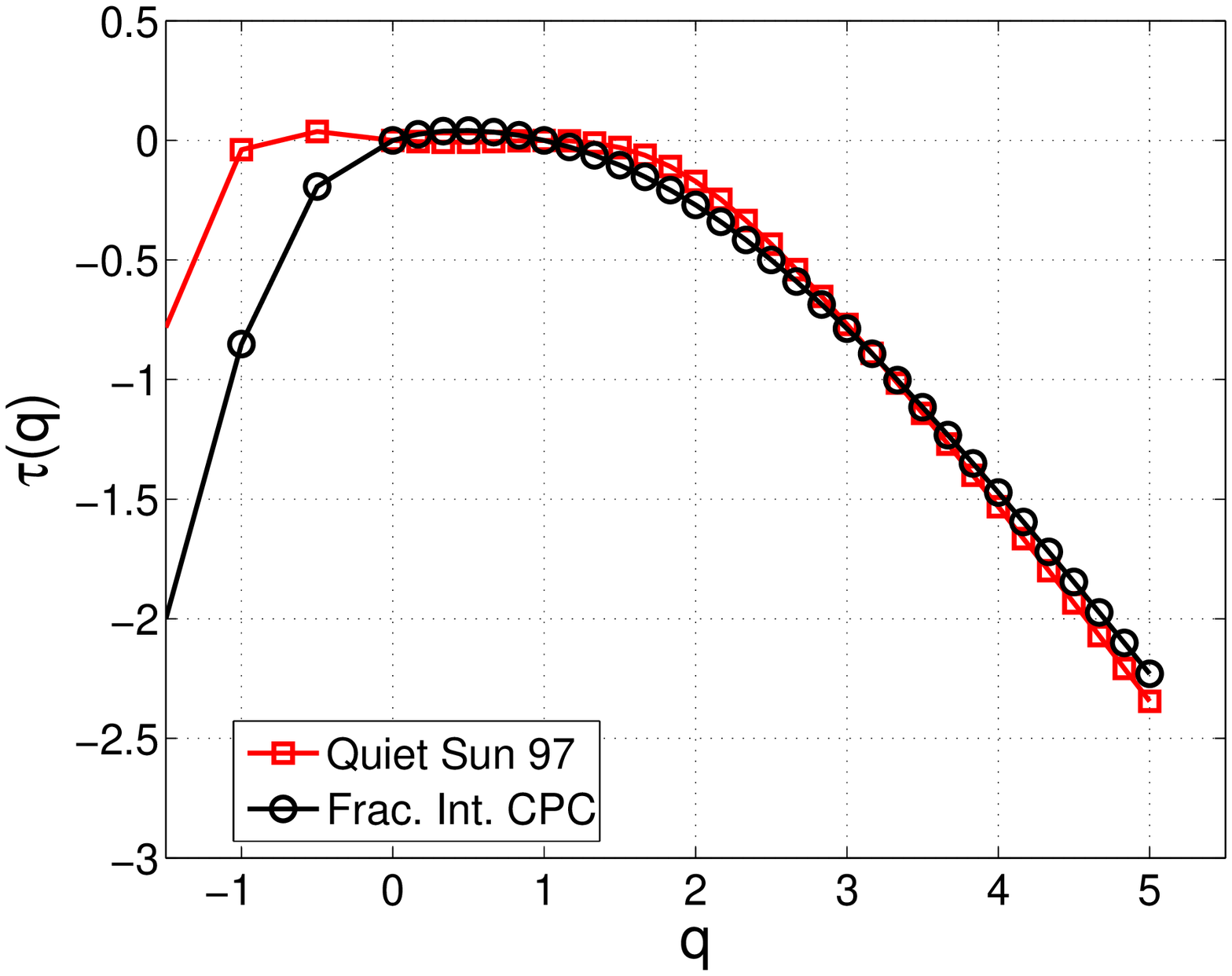}\\
	(a) & (b) & (c)
\end{tabular}
\end{center}
\caption{\label{figMFanal} {\bf (a)} 
Structure function ($q=-0.5, 1, 2,$ and $3$) estimated from DWT of Quiet Sun images in 1997. 
{\bf (b)} Exponents $\zeta(q)$ obtained from the DWT structure functions, for the original Quiet Sun images, and for simulated images using Compound Poisson Cascades
{\bf (c)} Exponents $\tau(q)$ deduced from $\zeta(q)$ by $\tau(q)=\zeta(q)-q\zeta(1)$.
}
\end{figure*}

We estimated the structure function $S_{DWT}(q,j)$ as in Equation (\ref{strucfun}) using the 54 Quiet Sun images that compose our data set. Figure~\ref{figMFanal}(a) shows the scaling behaviors of $S_{DWT}(q,\cdot)$ for  some values of $q$. A linear behavior of $\log_2 S_{DWT}(q,j)$ as a function of $j=\log_2 a$ is observed for $2\leq j\leq 5$. This is indicative of a scale-invariant behavior of the set of EIT images in this scale range.

The exponents $\zeta(q)$ are estimated from a linear regression
performed on $\log_2 S_{DWT}(q,j)$ in the scaling range $2\leq j \leq 5$. 
Figure~\ref{figMFanal}(b) shows the set of the resulting $\zeta(q)$ estimates for $-1\leq q\leq 5$. Note that errorbars are computed as the empirical standard deviation of the set of 54 estimates. As remarked above, estimates for negative values of $q$ are numerically unstable; this yields a large variability that results in wide errorbars. 
 Error bars for $q\geq 3$  becomes important as well. We will discuss this point in section~\ref{S:structure}.
 
 We observe a flattening of $\log_2 S_{DWT}(q,j)$ at small scales (for $j=1,j=2$). It may be due to the PSF that is larger than the pixel size, or to the fact that the wavelet coefficients capture mostly noise at the finest scale. 
It is interesting to note another flattening of the curves for $j\geq 5$ (and even a maximum for $S_{DWT}(3,j)$). There, scale invariance breaks down and we observe a characteristic scale of approximately 64 pixel wide ($115\mbox{Mm}$ on the Sun). Typically, such characteristic scale appearing in a structure function is indicative of the scale of injection of energy in the turbulent flows. In our case, it is compatible with the order of magnitude of the size of super-granules, an important structure that  governs  the turbulence in the corona.

In Section~\ref{S:fromEIT} we will compare intensities between EIT and rebinned-EIT images. Our aim is to draw conclusions that will remain valid as long as the scale-invariance observed on EIT images continues to prevail at EIT sub-pixel scales. As such, we must limit the amount of coarsening so as to stay within the scale-invariance range observed on EIT images. With the characteristic scale appearing on Figure~\ref{figMFanal}(a), this means that the rebin must be done on blocks of size smaller than $32\times 32$.


\subsection{Validity and limitations of multifractal analysis}
\label{S:structure}

When considering estimates of moments of higher orders, one may wonder about 1/ their existence and 2/ their precision. We examine hereafter two arguments which show that the $\zeta(q)$ estimated from Quiet Sun images must be considered differently if $q\leq 2.25$ or $q\geq 2.25$.

The first argument \citep{tddw:high_order_mom} is based on the analysis of the integrand of the theoretical quantity 
\begin{equation}
\label{E:integralS}
S^T_{DWT}(q,j) = \int_{0}^{\infty} p(y_j) y_j^q dy_j
\end{equation}
where $\mathbf{y}_j = |d_{j,\cdot} |$ denote the set of wavelet coefficient modulus at scale $j$, and
$p({y}_j)$ is their probability density function (pdf).  Then $S_{DWT}$ in~\eqref{strucfun} is viewed as an empirical estimate of the \lq true' quantity $S^T_{DWT}$. 
\cite{tddw:high_order_mom} indicates how to estimate the maximal order of moment, $q_{\max}$, such that the integral in~\eqref{E:integralS} begins to diverge. 
When applied to our 54 EIT images (only the centered parts of size $256\times 256$), the method gives $q_{max} \simeq 2$ for the first two finest scales, and  $q_{max} \simeq 1$ for the last three coarse scales.

Note that the test proposed in~\cite{tddw:high_order_mom} is not a proof that the moment estimates do not exist for $q>q_{max}$, but rather an indication that the variance of these estimates is large. 
One can still numerically estimate the structure functions $S_{DWT}(q,j)$ for $q\geq 2$ and get some information from its scaling behavior. 
Indeed, theoretical results on multifractal analysis~(\cite{baman02}) shows that the multifractal formalism (see Appendix~\ref{S:legendre})  remains valid for $h_+^*\leq h\leq h_-^*$ such that $D(h)\geq 0$ only, i.e.~for $q_-^*\leq q\leq q_+^*$. In two dimensions,    $q_+^*$  is found by solving the equation;  $\zeta(q)=2-q\zeta'(q)$, where $\zeta'(q)$ denotes the derivative of $\zeta(q)$. Moreover, for $q\geq q_+^*$, one observes the \emph{linear behavior} $\zeta(q)=2-q\zeta'(q_+^*)$ described in~\cite{lac04}.

Since our aim is to propose a multifractal model of Quiet Sun images, this model must obey the same linearization effect as the EIT images, hence the importance of estimating $q_+^*$. 
The arguments in~\cite{tddw:high_order_mom} and in~\cite{lac04} are consistent in showing that a lack of statistics (for large $q$) will induce a growing variance of the estimate, but that the potential bias will remain limited.
To summarize, we will perform a multifractal analysis of Quiet Sun images for $0\leq q\leq 5$ because our  estimates are numerically unstable for $q<0$, and because we need to identify the critical order $q_+^*$. For Quiet Sun images, we obtain the value of $q_+^*\simeq 2.25$.

\section{Synthesis of Quiet Sun-like images}
\label{S:synthe}

\subsection{Physical interpretation of multiscaling exponents}

When considering the data as a multifractal field, it is usually seen as based on some underlying positive multifractal scalar field. For instance, let us recall that the Kolmogorov 1962 (K62) theory of turbulence \citep{K62} proposes to describe the statistics of the velocity increments at scale $r$ by $\langle |\delta_r v|^q\rangle\sim \langle\varepsilon_r^q\rangle r^{q/3}$ where $\langle \varepsilon_r \rangle$ stands for the locally averaged energy dissipation at scale $r$. Then, Kolmogorov postulates that the dissipation is intermittent and obeys scale invariance law so that $\langle\varepsilon_r^q\rangle\sim r^{\tau(q)}$. With this definition, $\tau(0)=0$. Moreover, in an homogeneous system, the spatial mean at scale $r$, $\langle\varepsilon_r\rangle$, does not depend on $r$, and hence $\tau(1)=0$. As a consequence, K62 theory predicts that $\langle |\delta_r v|^q\rangle\sim r^{q/3+\tau(q)}$. The link between the scaling behavior of the dissipation $\varepsilon_r$ and that of the velocity field is expressed in terms of the velocity increments $\delta_r v$. Thus, it suggests to look at the velocity field itself as a fractionally integrated version of order $H=1/3$ of the dissipation field. In the Fourier domain, this translates into a $1/\|\k\|^{H}$ filtering (with $H=1/3$ in turbulence). From a mathematical point of view, the dissipation field would be modeled by the density of a multifractal measure (or by a probability density function of some distribution).

In Figure~\ref{figMFanal}(b)  we see that 
 for the intensity of Quiet  Sun images we get 
$\zeta(1)=0.55 \pm 0.06$. This indicates that the underlying signal is rather similar to the velocity field than to the dissipation field above. Therefore, the intensity field of Quiet Sun images cannot be directly modeled by the density of some multifractal measure while these measures are the easiest multifractal objects to built and simulate. To overcome this apparent difficulty, we propose to model EIT images by a (fractionally) integrated version of some multifractal  density characterized by a set of multifractal exponents $\tau(q)$. The multiscaling exponents of the resulting intensity field will be such that 
\begin{equation}
\label{zetaq}
	\zeta(q)=qH+\tau(q)~,
\end{equation}
where $H=\zeta(1)$ represents the \lq fractional order' of the integration process. Fractional integration here  loosely means a $1/\|\k\|^{H}$ low-pass filtering, where $\k$ is the Fourier frequency.

\subsection{Fractionally integrated Compound Poisson cascades}
\label{CPC}

Multifractal measures are typically generated by means of a multiplicative cascade process. A measure is initially distributed uniformly over a set.  An iterative division of the measure among subsets, and next sub-subsets is then performed up to infinity. The division is done according to an invariant allocation rule, which in the most interesting examples is probabilistic. 
Several authors have introduced precise definitions of multifractal measures in one dimension~\citep{sm01,baman02,muba02,bamu03,cra03a,cra05ieee,s03}, and recently in dimension $D\geq 2$ \citep{c06,c07pami,sc07} for image modeling purpose mainly. 
An important subset of the family of multiplicative cascade processes is that of Compound Poisson Cascades (CPC). 
CPC were originally introduced by~\citet{baman02}. They provide us with a model to generate multifractal densities with prescribed multiscaling exponents $\tau(q)$, and moreover their numerical synthesis is easy.
CPC allows to generate the necessary underlying multifractal density evoked above. Their role with respect to the intensity in EIT images is equivalent to the role of the dissipation field $\varepsilon$ with respect to velocity increments in K62 theory of turbulence.
See Appendix~\ref{A:multifrac} for a more detailed presentation of Compound Poisson Cascades and their numerical synthesis.

The main ingredient that controls the multifractal exponents $\tau(q)$ of a CPC is the distribution of the so-called multipliers denoted by $W$. Indeed, once the law of the $W$ is determined, one has $\tau(q)=q(\E W-1)+1-\E W^q$ 
($\E$ denotes mathematical expectation) for $q_-^*\leq q\leq q_+^*$ (see Section~\ref{S:structure} and Appendix~\ref{A:multifrac}). As a consequence, the main thrust here is to propose a modeling of Quiet Sun images by fractionally integrated Compound Poisson Cascades.  In practice the fractional integration corresponds to a $1/\|\k\|^H$ filtering which is carried out in the Fourier space thanks to a fast Fourier transform and by using a $1/\|\k\|^H$ frequential response that is truncated near the origin $\k=0$ (since it is not defined at this point). We have chosen to impose a saturation at the value associated to the smallest available discretized frequency.This low-pass filter precisely modifies the multifractal exponents so that given $\tau(q)$ one gets a multifractal intensity-like field with multifractal exponents
\begin{equation}
\label{ztheogeneral}
	\zeta_{CPC}(q)=
	\left\{
		\begin{array}{lcl}
			\ds{qH+\tau(q)} & \mbox{ for } & 0\leq q\le q_+^*\\
			2-q\zeta'(q_+^*) & \mbox{ for } & q\ge q_+^*~,		
		\end{array}
	\right.
\end{equation}
where $\zeta'$ denotes the derivative of $\zeta$.
The next section explains how to optimize the fit of a CPC stochastic process to the modeling of Quiet Sun images.

\subsection{Model identification}
\label{S:identif}
The model identification concerns two quantities : the fractional order $H$ of integration, and the multiscaling exponents $\tau(q)$ in Equation~(\ref{ztheogeneral}). The parameter $H$ describes the linear trends of $\zeta(q)$. The function $\tau(q)$ is a non linear concave function obeying $\tau(0)=\tau(1)=0$ and controls the multifractal behavior of the final process. 
As a consequence, one expects that $H=\zeta(1)$ so that we will use the estimated $\zeta(1)$ as an estimate of $H$. We get $H=0.55\pm 0.06$. Next, $\tau(q)$ is estimated from the relation $\tau(q)=\zeta(q)-q\zeta(1)$. 

We studied the  adequateness of several compound Poisson cascade models to our experimental data. The best fit was obtained with a model such that  $W = ((1+T)^{1/T}u)^T$ where $u$ is uniformly distributed in $[0, 1]$. Such a model is characterized by
\begin{equation}
\label{E:taucpc}
	\tau_{CPC}(q)=
		1-\frac{(1+T)^q}{(1+qT)} \quad (\mbox{for } 0\leq q\le q_+^*)
\end{equation}
with $T=0.85$ (see \cite{c06,c07pami} for a detailed presentation of available models).
The existence of an upper bound $q_+^*$ originates from  the linearization effect described in~\citep{lac04}, cfr Section~\ref{S:structure}. Therefore, one expects $\zeta(q)$ to behave as
\begin{equation}
\label{ztheoCPC}
	\zeta_{CPC}(q)=
	\left\{
		\begin{array}{lcl}
			\ds{qH+1-\frac{(1+T)^q}{(1+qT)}} & \mbox{ for } & 0\leq q\le q_+^*,\\
			2-q\zeta'(q_+^*) & \mbox{ for } & q\ge q_+^*,		
		\end{array}
	\right.
\end{equation}
where the best fit is obtained with $q_+^*\simeq 2.25$ and $T=0.85$. The estimates obtained from the multifractal analysis of 54 realizations are shown on figure~\ref{figMFanal}(b) (black circles): they are quite consistent with the theoretical values, except for $q\leq 0$ that we do not consider here because of numerical instability.

\subsection{Validation  of the model}

Once we have identified the parameters of the model described above, we can numerically synthesize as many realizations as needed, at any desired resolution. 
To test the validity of our model, we consider several indicators: the exponents $\zeta(q)$ and $\tau(q)$, the Fourier spectrum, the histogram, and the visual aspect of synthetic images.

First, we generate  54 independent  realizations of 512$\times$512 model images, and we estimate the exponents  $\zeta(q)$ and $\tau(q)$. We then compare these estimates with the values of  $\zeta(q)$ and $\tau(q)$ obtained in Section~\ref{QSMFA}, Figure~\ref{figMFanal}(b)-(c) on  54 EIT Quiet Sun images.
 Figure~\ref{figMFanal} shows that estimates on the images from the CPC model are nearly superimposed onto estimates from Quiet Sun images.

Second, note that by construction, the Fourier spectrum of the 2D multifractal density underlying our synthetic image process is $\sim 1/k^{2+\tau(2)}$ where $\tau(2)<0$ so that $2+\tau(2)<2$~\citep{chainaistpami07}.
As a consequence, the spectrum of our fractionally integrated model is $\sim 1/k^{2+\tau(2)+2H}$. Considering all together the 54 simulated images of size $512\times 512$, the slope of their omnidirectional Fourier spectrum is equal to $-2.8$ which is identical to the value of the slope computed on the EIT data set. 
 
Third, we compare in
Figure~\ref{F:histoIlogI97_powerspect}(a)
the histograms of
 fractionally integrated compound Poisson cascades, of EIT images, and of a lognormal fit on the EIT data set. The three histograms have the same mean and variance, and are computed on the same number of realizations.

 Both EIT and fractionally integrated CPC histograms exhibit a slow decrease for high intensity values, albeit with a different slope.  The CPC process attributes too much weight to small values, and not enough to large values, while the bulk of the distribution is fairly similar in both cases. 
 Fitting 19.5nm EIT intensities with a single distribution has been proved difficult, see e.g.~\citet{2000ApJ...544..550A}.  
 Ideally we should constrain 
  the histogram of our simulated CPC images to be the same as those of EIT images. 
 However building such a stochastic process with prescribed multifractal spectrum \emph{and}  histogram is an 
    intricate problem.

 Indeed, multifractal properties can be interpreted as joint properties on the histograms of the image seen at different resolutions. However, not any distribution of probability is compatible with some given multifractal properties. Trying to naively impose such an histogram at some given resolution in the model (by simply rescaling the data)  disturbs and even kills the multifractal properties. While it is easy to adapt the histogram of an image to some desired shape, it is much more difficult to impose scale invariance properties. In the present work, we are particularly interested in the extrapolation of the properties of images at smaller resolution. Therefore we have chosen to focus on the evolution of the probability density functions through the scale rather than on the specification of the histogram at some particular scale.  In Section~\ref{S:conclusion}, we outline some avenues for succeeding in  preserving both the histogram and the multifractal spectrum.

\begin{figure*}
   \begin{center}
     \begin{tabular}{cc}
          \includegraphics[height=55mm]{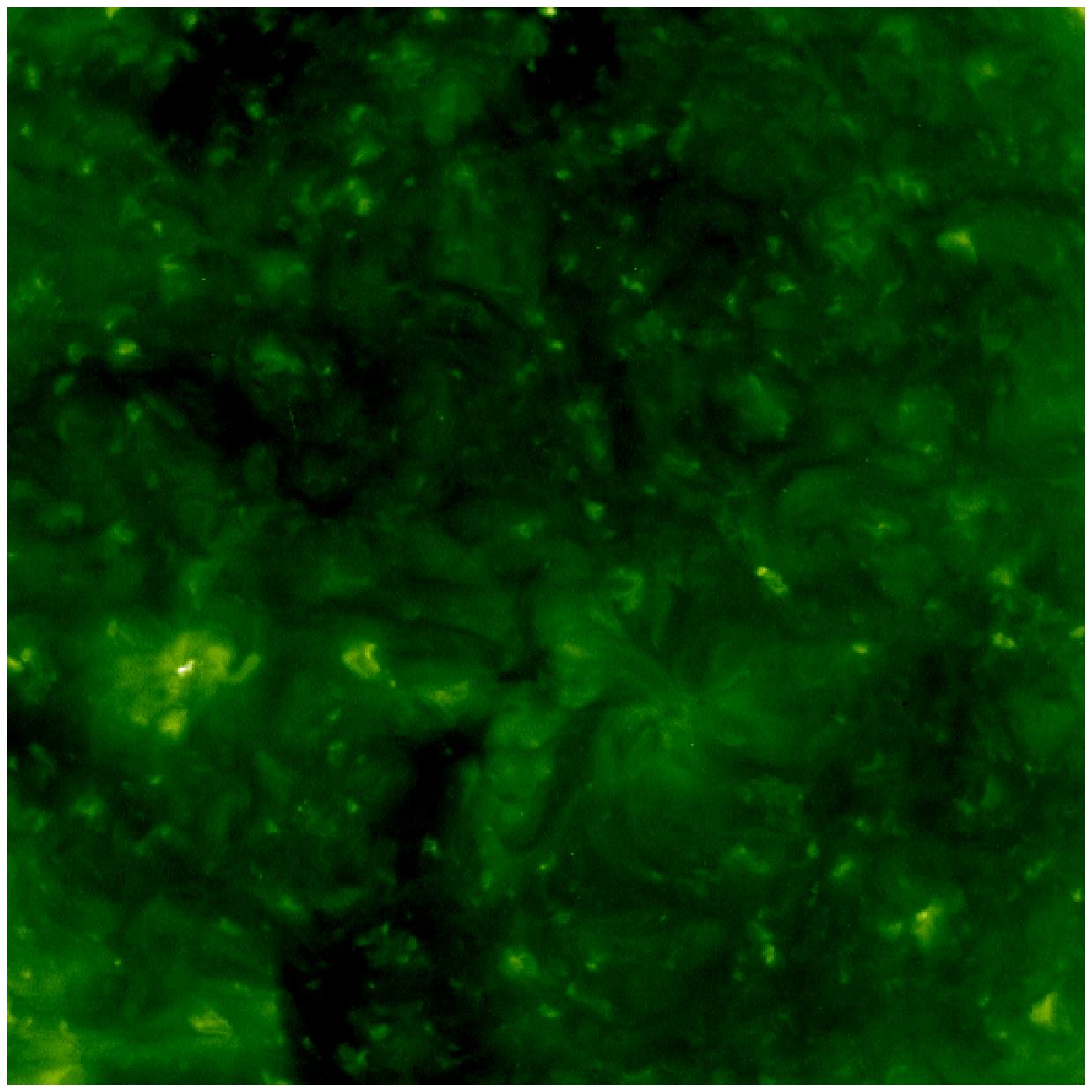}
       &\includegraphics[height=55mm,width=55mm]{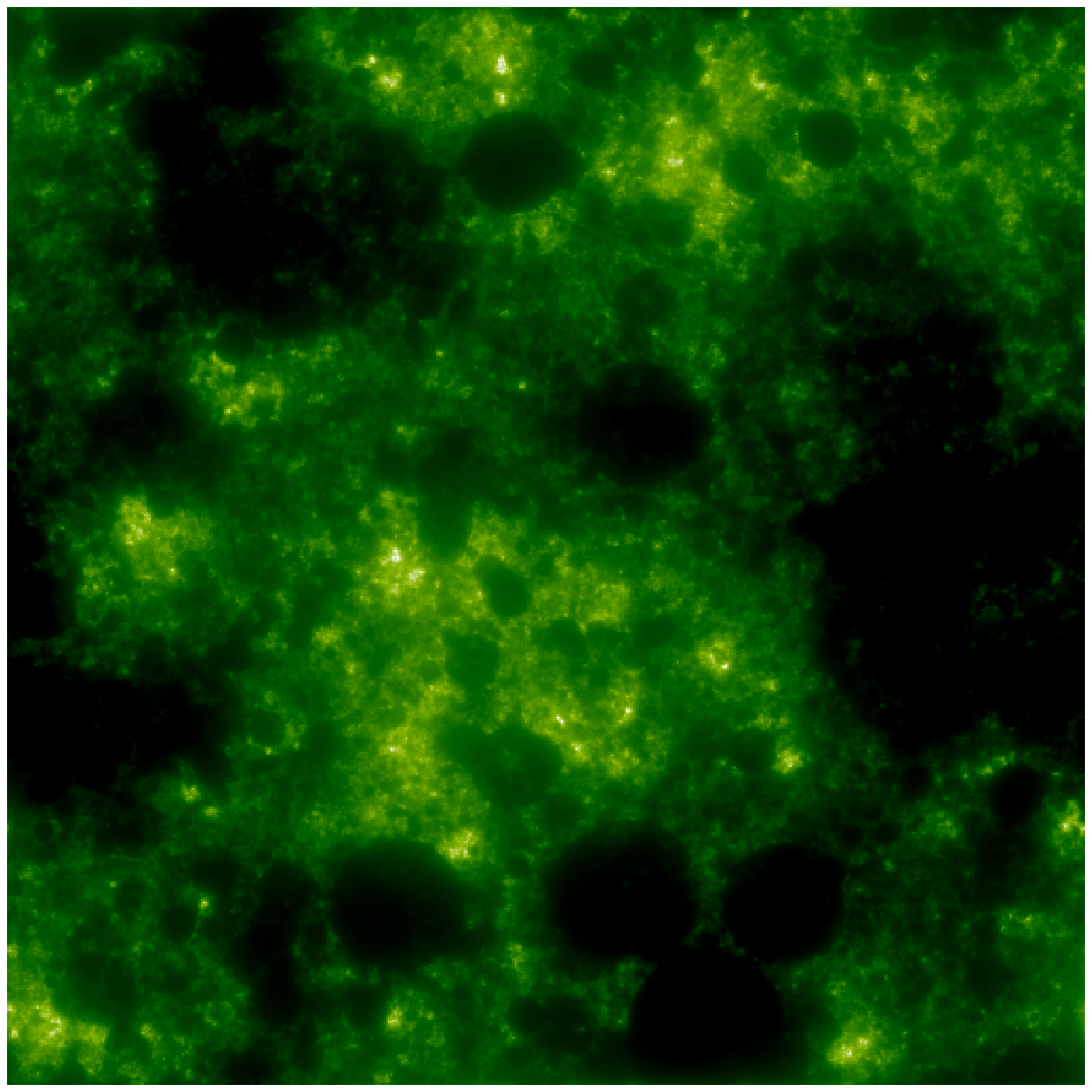}\\
        (a) & (b) 
     \end{tabular}
   \caption{\label{fig1} Examples
     of 512$\times$512 images of (a) Quiet Sun, (b)
     a fractionally integrated compound Poisson cascade.}
   \end{center}
 \end{figure*}

Finally, Figure~\ref{fig1} allows to compare visually an EIT image together with a realization from a fractionally integrated CPC. The visual aspect is particularly sensitive to high intensity values. Note that Figure~\ref{fig1}(b) was obtained directly from our synthesis model without any post-processing of the image.

\begin{description}
\item[{\bf Remark}] {Figure~\ref{figMFanal} represents both the $\zeta(q)$ exponents and the $\tau(q)$ exponents (where the linear trend of the $\zeta(q)$ has been removed) to make comparisons more discriminating. Errorbars reflect the empirical standard deviation of the set of 54 estimates. They are not computed as an estimate of the variance of the estimator for one single image.} 
\end{description}

In the previous sections, we studied the scale-invariance properties of QS images, and we proposed a stochastic modeling of these images based on compound Poisson cascades. Having evaluated the validity of our model with different indicators, it appears that the multifractal modeling of QS images can be  used as a benchmark when comparing intensities at different scales of observations. Such comparison is precisely the aim of the next section.

\section{How many pixels from HRI will have a good SNR?}
\label{S:fromEIT}

Two successive increases in resolution are taking place when comparing EIT at the Lagrange point $L1$ and HRI at perihelion: the first one is due to an enhanced pixel angular resolution, the second is due to a smaller distance to the Sun.  
One needs to answer the question:
\lq what proportion of radiant intensity remains available per pixel after such a magnification ?'

Equation~(\ref{E:SNRineq2}) gives us information about the SNR available given some experimental conditions, and it is straightforward to deduce from this equation the radiant intensity for a uniform source (Section~\ref{S:uniform}), and for a point-like source (Section~\ref{S:dirac}). However, the quiet corona does not behave as a uniform source, nor as a collection of infinitely localized sources.

Instead, we saw in Section~\ref{S:waveletMF}  that Quiet Sun images exhibit multifractal properties. Therefore, they are highly irregular and  can be seen as a superposition of  singularities with finite H\"older exponent. This situation is  intermediate between a uniform and a dirac signal and we show in Section~\ref{S:MFsignal} what it implies in terms of quality of the observations at high resolution.

We now introduce some notations. Let $L({s})$ be the radiance emitted from an elementary surface of the Sun  ${ds}$, and let $A_{sun}$ be the area at the Sun corresponding to one pixel.
The~\emph{radiant intensity} received by one pixel can be computed as $P = \int_{A_{Sun}} L(s)ds$. Recall that we are interested in the evolution of $Q\doteq P/d^2$, cfr. Equation~\eqref{E:eqQ} (the sign $\doteq$ means \lq by definition'). In this section, the subscript $LR$ (resp. $HR$) denotes a quantity at low (resp. high) resolution. Moreover, we assume that the PSF of the instrument adapts exactly to the decrease in pixel size.

\subsection{Uniform distribution of intensity}
\label{S:uniform}
When the angular resolution is increased by a factor $N$,  $A_{sun}$ is divided by a factor $N^2$, and hence
\begin{equation}
\label{E:P_HRhomo}
P_{HR} = \int_{A_{sun}/N^2}L(s)ds = \frac{P_{LR}}{N^2}~.
\end{equation}
Suppose next that the distance $d$ and the effective area $A_{eff}$ stay constant. The SNR of the high resolution telescope is penalized by a factor $N$ since
\[
{SNR_{HR}^2} = \frac{SNR^2_{LR}}{N^2}
\]

Consider now a telescope with a given resolution that gets closer to the Sun.
As the distance $d$ decreases, so does the area sustained by one pixel,  hence $A_{sun}$ decreases in the same proportion.
The radiance being independent on the distance, the SNR of a  uniform source does not change as long as $A_{eff}$ stays the same.

\subsection{Point-like source}
\label{S:dirac}
For a constant distance $d$, an angular resolution increased by a factor $N$, and a  signal containing one point-like structure, 
 if both the low and high-resolution pixel contains the same dirac $\delta$ located at $s_0$, we obtain:
\begin{equation}
\label{E:P_HRdirac}
P_{HR} =\int_{A_{sun}/N^2} L(s)\delta(s-s_0)ds =L(s_0)~,
\end{equation}
 which is invariant with respect to the factor $N$, hence $SNR_{LR}=SNR_{HR}$.
 
 If now a given telescope gets closer to the Sun, a pixel  that contains a dirac at $s_0$ will receive
\[
Q = \frac{L(s_0)}{d^2}
\]
and hence the SNR increases when $d$ decreases.

Table~\ref{T:snr_evol} summarizes these well-known situations. Note that the study of the temporal variability would produce an equivalent Table, showing the  difference between a continuous, regular evolution and a highly intermittent one. 
Table~\ref{T:snr_evol} suggests that 1/ observations will always get better when getting closer,
and 2/ regions where the source is uniform  will not benefit from an increase in resolution (SNR will be penalized by a $1/N$ factor), but irregular source regions will benefit.

\subsection{Multifractal signal}
\label{S:MFsignal}

We use our data set of 54 images of Quiet sun described in Section~\ref{S:dataset} as well as 54 images of size $512\times 512$ simulated with the fractionally integrated CPC process detailed in Section~\ref{S:synthe}. For all these images we keep only the central part of size $500\times 500$ so as to ease the computations below.

Our aim is to determine the proportion of pixels having at high resolution a SNR larger or equal to the SNR available at low resolution, and this for a given increase in effective area $A_{eff}$.  
These pixels will be qualified as \lq SNR-preserved'.
Recall from Equations~(\ref{E:SNRineq2})-(\ref{E:eqQ}) that 
\begin{equation}
\label{E:recallSNR}
SNR^2 \leq A_{eff}\frac{P}{d^2} \doteq A_{eff}Q~,
\end{equation}
where $P$ is the radiant intensity. The inequality $ A_{eff,HR}Q_{HR} \geq A_{eff,LR}Q_{LR}$ is verified as soon as
\begin{equation}
\label{E:gainAeff}
G_{Aeff} \doteq \frac{A_{eff,HR}}{A_{eff,LR}} \geq \frac{Q_{LR}}{Q_{HR}}~,
\end{equation}
where $G_{Aeff}$ denotes the gain in effective area. 
Since the solar corona is not homogenous, Equation~(\ref{E:gainAeff}) is pixel-dependent.
$Q_{HR}$ is given by the intensity in  EIT images (real or artificial). In order to obtain $Q_{LR}$, we start from these images and we derive an appropriate rebinned version that simulates a telescope having a smaller angular resolution, or a space-craft located at a larger distance from the Sun. Naturally, the transition from high to low resolution must match the evolution summarized in Table~\ref{T:snr_evol}. 

The study of the ratio $Q_{LR}/Q_{HR}$ (and hence of a lower bound for $G_{Aeff}$)
 finds its justification under the self-similarity assumption: if the scale-invariance observed on EIT images continues to prevail at sub-pixel scales, the ratio $G_{A_{eff}}$ will have a similar distribution
as soon as  the difference in resolution considered is kept fixed. 

Let $I_h$ denote a high-resolution image. We term the pixel in this image \emph{micro-pixel}. As the spatial resolution decreases by a factor $M^2$, a  low resolution image $I_l$ is created, for which each pixel, called \emph{macro-pixel}, contains $M^2$ \emph{micro-pixels} from $I_h$. We  assume that the macro-pixels are disjoint. 
We analyze what happens with 1/ a change in angular resolution 2/ a change in distance to the Sun 3/  a combination of both effects.

\subsubsection{Change in resolution}
\label{S:change_res}

From equation~(\ref{E:recallSNR}), we see that if $d$ and $A_{eff}$ are constant, a change in angular resolution will impact the SNR through the radiant intensity $P$ only. Since $P$ is proportional to the area observed on the Sun, the total intensity observed on a given area at the Sun must be preserved.

We start from an image (real or simulated) at EIT resolution, that stands as \lq high-resolution' image. We then construct a low-resolution image by rebining blocks of size $N\times N$ such that the intensity in a macro-pixel is equal to the sum of $N^2$ corresponding micro-pixels

A magnification $N=5$ corresponds to the change in pixel angular size between EIT and HRI or TRACE;  indeed EIT  has a pixel angular size of $2.6$~arcsec, whereas the one of HRI and TRACE is of $0.5$~arcsec.
We are interested in the increase in effective area needed to preserve  the SNR at high resolution.  The ratio $G_{Aeff}$ between the intensity at low resolution (in a macro-pixel) and at high resolution (micro-pixel) precisely gives this wanted factor  coefficient for $A_{eff}$.

\begin{figure*}[t]
\vspace*{2mm}
\begin{center} 
  \begin{tabular}{cc}
\includegraphics[width=8.3cm]{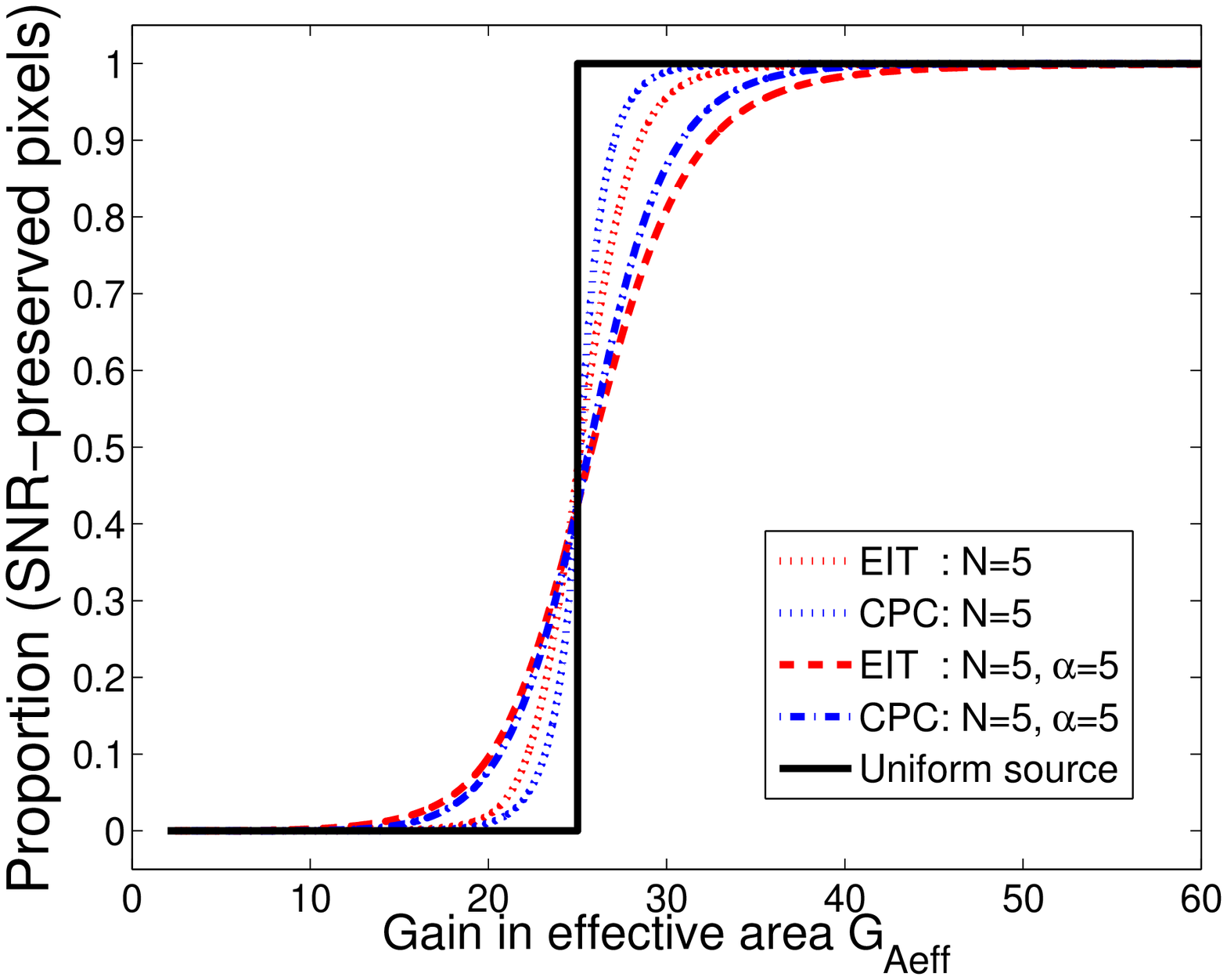} 
&\includegraphics[width=8.3cm]{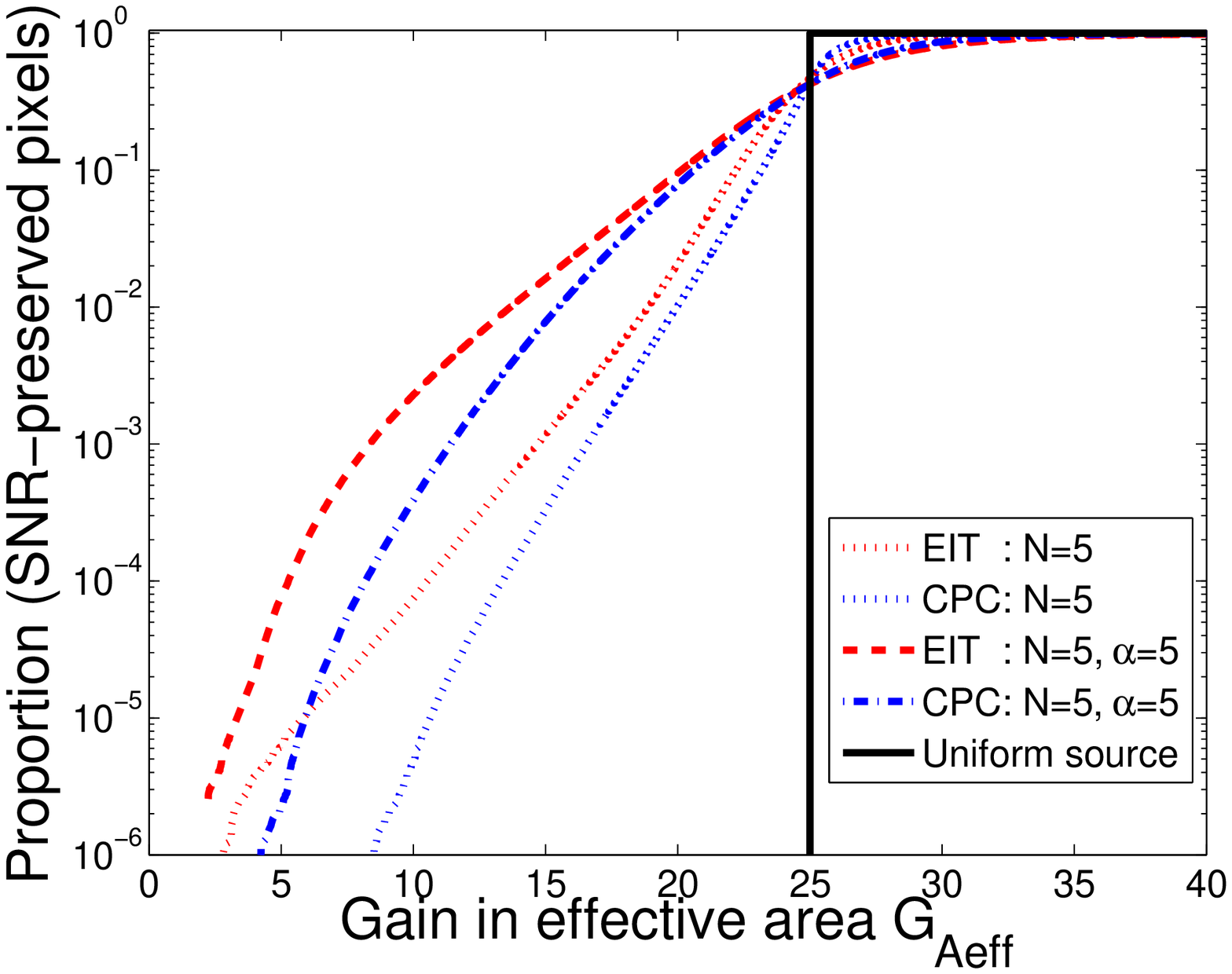}\\
        (a) & (b)  
  \end{tabular}{cc}
   \caption{\label{F:ratiolowNlarged}  Simulation of the
   proportion of pixels in an HRI instrument at perihelion that will conserve the same SNR  as in the case of EIT at L1 (we term these \lq SNR-preserved pixels') 
   given a factor of increase of the effective area in (a) linear and (b) semi-logarithmic representation. Both EIT data (Real) and synthetic data (CPC) are represented in case of 1/ a better angular resolution ($N=5$), 2/ a better angular resolution and a smaller distance ($N=5,\alpha=5$). For a gain in effective area equal to $15$, $1.6\%$ of the pixels are SNR-preserved on real EIT data when $N=5,\alpha=5$. This figure becomes $0.2\%$ when $G_{Aeff}=10$ only.}
   \end{center}
 \end{figure*}

Figure~\ref{F:ratiolowNlarged} represents, for a given value of the gain in effective area $G_{Aeff}$, the percentage of pixels at high resolution that are SNR-preserved. The dotted lines correspond to the case where only the factor $N$ changes.

If a new telescope has $N=5$ better a resolution and 15 times larger effective area ($G_{Aeff}=15$), then the simulation on EIT data shows that out of  $1024\times 1024$ pixels about 1000 pixels are SNR-preserved; this number becomes 300 when the artificial CPC images are used.
When $G_{Aeff} =25$ almost all pixels keep the same SNR as in the low-resolution image.

\begin{figure*}[t]
\vspace*{2mm}
\begin{center}
\includegraphics[width=8.3cm]{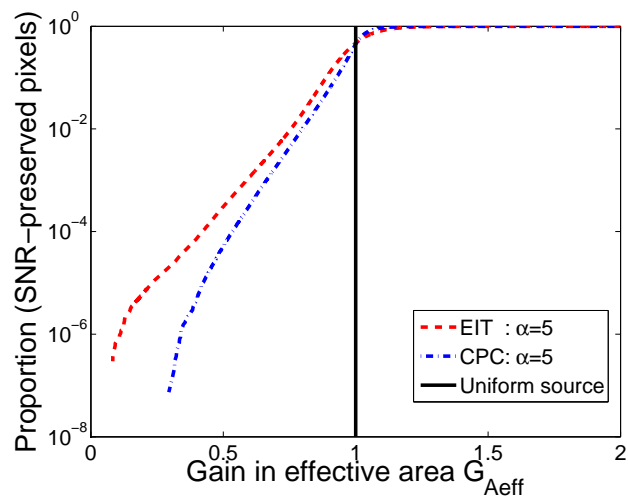}
   \caption{\label{F:ratiolarged}  
   Proportion of pixels having a similar SNR at low resolution (far from the Sun) and at high resolution ($\alpha=5$ times closer to the Sun) for a given change in effective area. When getting closer to the Sun, one may have a reduced effective area and still have a proportion of pixels with preserved SNR.}
   \end{center}
 \end{figure*}

Note that in the EIT data set there were  obvious cosmic rays, that produce a value of  ${G_{Aeff}}$ as small as 1.4. The graphs in Figures~\ref{F:ratiolowNlarged} and~\ref{F:ratiolarged} represent the results without these outliers.

\subsubsection{Change in distance}
\label{S:change_dist}

From Sections~\ref{S:uniform} and~\ref{S:dirac} we know that when getting closer to the Sun, the SNR increases in case of dirac-like signals, and remains constant for a flat source. 
In other words, the  \emph{mean} value of $Q$ over a given area of the Sun  stays constant as the distance Spacecraft-Sun changes.
This means that for some pixels 
the ratio $G_{A_{eff}}$ between intensity at low (far from the Sun) and at high (closer to the Sun) resolution may be smaller than one, i.e. we may keep constant or even decrease the effective area, and still have $SNR_{HR} \geq SNR_{LR}$.

Let $\alpha = \frac{d_1}{d_2}$, where $d_1$ (resp. $d_2$) is the distance \lq far from the Sun' (resp. \lq close to the Sun').
Solar Orbiter will get as close as $0.22$~AU at perihelion; this is about five times closer to the Sun than current telescopes, we therefore illustrate with Figure~\ref{F:ratiolarged} the case where $\alpha=5$. Note that a closer proximity to the Sun may imply to build a telescope with a smaller pupil diameter, and hence a reduced radiant intensity. However, we do not discuss this aspect here.

\subsubsection{Change in distance and in resolution}

We now simulate a change in distance to the Sun and in resolution by two successive rebins.
With the first rebin  on blocks of size $5\times 5$, the average pixel intensity is preserved ( modeling thus a larger distance to the Sun). In a second rebin of the same size $5 \times 5$, the sum intensity is kept constant within a pixel (simulating a coarser-resolution telescope).
Note that a rebin in one dimension of a factor 25 still falls within the scale invariance range shown in Figure~\ref{figMFanal}. Starting from images of size $500\times 500$, low-resolution images of size $20 \times 20$ are constructed.

If we bring the new telescope (for which $N=5, G_{Aeff}=15$) five times closer to the Sun, then according to the simulations using EIT images, $1.6\%$ of the pixels are SNR-preserved. This figure becomes $0.9\%$ when artificial data based on CPC model are used.
Figures~\ref{F:ratiolowNlarged}(b) and~\ref{F:ratiolarged} show that for a given $G_{Aeff}$, the proportion of pixels that are SNR-preserved is underestimated by artificial CPC images.  This follows from the fact that
 the CPC model does not put enough weight on large intensity values (often featuring bright points, i.e.~dirac-like structure) as compared to EIT images, cfr. the discrepancy observed between EIT and CPC histograms for high intensity values.

\subsection{Implications for new High Resolution telescope}
Our study has several implications for the conceptions of high resolution EUV telescope. 

In a very high-resolution EUV telescope with fixed exposure time, pixels with low photon counts may contain more noise than signal.   In this case, one would need compression algorithms that adapt to  the level of signals, viz.~that compress more where the SNR is low.
  
In order to decrease the readout noise in the image, one possibility is to allow for multiple non-destructive readouts (NDR) and combine successive NDR values~\citep{finger:MultiNDR}.
 
 To have a larger  dynamical range, one could allow for exposure times $T$ that vary according to the photon flux in each pixel: the less the photon flux $\phi(t)$, the longer $T$ so that $\int_T \phi(t)dt$ would ideally be constant for all pixels. The map of these exposure times $T$ would be the output of such a device~\citep{patent:exptime}. 
 
 To motivate such options the radiometric models  will evaluate for a given signal the photon counts expected. It is important, especially for high resolution EUV telescopes to have an accurate model; and towards this goal it is necessary to take into account the multifractal nature of the data. In this respect, the stochastic model proposed here provides a natural benchmark and a correction factor. 
 
 In the 17.1~nm passband, the effective area of EIT is of the order $10^{-2}cm^2$. New technologies will allow an HRI instrument to have an effective area close to $10^{-1}cm^2$, so that between HRI and EIT we would have a gain in effective area equal to ten approximately.     
Our study shows that with $G_{A_{eff}} =10$ about $0.2\%$ of the pixels in a HRI QS image will have a SNR similar to what is observed on EIT images. If  $G_{A_{eff}}$ is increased up to 15, $1.6\%$ of the pixels will be SNR-preserved, i.e. about eight times more pixels will have an excellent SNR. This highlights that a relatively small increase in instrumental performance may have a large impact on the quality of the data. 

\section{Conclusion and future prospects}
\label{S:conclusion}

Amongst the many challenges faced by high resolution missions, the precise estimation of photon counts is of fundamental importance. We showed in this paper that an accurate radiometric model must take into account the spatial inhomogeneities present in the source.

We first showed how to characterize these spatial inhomogeneities through a multifracal analysis of Quiet Sun images:
We computed the multifractal spectrum and derived a  set of parameters, namely the multiscaling exponents, that quantitatively describe the scale invariance properties. We then proposed a family of stochastic processes that obey the same statistical property with the same set of parameters, thus injecting minimum a priori in the model. This establishes firmly the spatial scale-invariance structure of EIT images.

 Next, we compared a set of images (real and simulated
 via the above stochastic process) at EIT-resolution with
an appropriate rebin of these images. The rebin is done so as to simulate a larger angular resolution, and a larger distance to the Sun. By comparing the intensity values at low and high resolutions, it is possible to estimate the needed gain in effective area such that the SNR at high and low resolutions remains the same. 
  If the scale-invariance observed  at supra-pixel scales
on EIT images continues to prevail at
smaller scales, the distribution of this ratio will be the same when comparing 
 current low resolution EIT images with 
 future high resolution 
images. In this sense, we  provided a proxy for the needed gain in effective area for HRI telescopes.

With the new technologies, the future HRI instrument will have an effective area which is approximately 10 times larger than the one of EIT. In this case, our study shows that about $0.2\%$ of the pixels in a HRI QS image will have a SNR similar to what is observed on EIT images. However, if the effective area of HRI becomes 15 times larger (instead of 10 times), about $1.6 \%$ of the pixels (i.e.~eight times more pixels) will have an excellent SNR.  Hence a relatively small increase in instrumental performance may have a large impact on the quality of the data. 

The results presented in this paper are based on images of the warm corona. X-ray images of the hot corona are likely to exhibit a more pronounced spatial intermittency;  a similar analysis on X-ray images should therefore conclude that  a smaller gain in effective area is needed. 
As a general conclusion we can say that 1/ all other things being equal, the quality of observations will improve when getting closer to the Sun\footnote{However, as said before, a closer distance to the Sun might imply to build a telescope with a smaller pupil, and hence a reduced radiant intensity.}; 2/ regions where the source
is uniform will be penalized by an increase in angular resolution, but irregular source regions will benefit from such an increase.

In order to help improve current radiometric model, our next goal is to generate synthetic images at arbitrary resolution (e.g.~EIT resolution or higher) which would capture as many statistical properties of Quiet Sun images as possible.
Indeed, Figure~\ref{figMFanal} shows that we have succeeded in building 
synthetic images from multifractal processes that obey some of the scale-invariance properties of Quiet Sun EUV images, more precisely that have the same multiscaling exponents.
However, in Figure~\ref{F:ratiolowNlarged} we see that the present model underestimates the proportion of bright (i.e. SNR-preserved) pixels.  This is related to  the discrepancy  between the histograms of real and artificial EIT images observed for high intensity values. Indeed, our proposed model
captures the self-similar and multifractal nature of Quiet Sun images, but does not take into account other constraint such as the distribution of intensities in EIT images.

In future work, we plan on exploring two alternatives to obtain a model that would both preserve the  histogram observed at a given resolution and satisfy the multifractal properties. The first one is to find a way to constrain the synthetic processes to comply with  some properties on their marginal distributions (like e.g. intensity
histograms) as well as with the desired scale invariance properties. 
 The difficulty here comes from the fact that the multifractal spectrum strongly constrains the global structure of stochastic process. 
Hence there are few degrees of freedom left to further adopt  other criteria. This issue remains thus an open question for mathematicians.
The second possibility, more empirical, is to build  higher resolution images from low resolution images. The challenge here is  to extrapolate e.g.~EIT images at higher resolution using the scale invariance properties. In this case, the real image would serve as a boundary condition of the model, and by construction the histogram of the low resolution image would be preserved.

\appendix

\section{Multifractal formalism and Legendre transform} 
\label{S:legendre}
Multifractal analysis aims at the characterization of the regularity of measures, functions or graphs of realizations of a stochastic process~(\cite{jaf97ab}).
Often, one quantifies the  presence of singularities in an image thanks to the multifractal spectrum $D(h)$, where $h$ is the so-called H\"older exponent.
 In brief, $f(\x)$ is said to be locally H\"older regular with exponent $h(\x_o)$ at $\x_o$ if $h(\x_o)$ is the highest exponent such that there exist a polynomial $P(\x)$ and a constant $C$ with $|f(\x)-P_{\x_o}(\x-\x_o)|\leq C|\x-\x_o|^h$ for $\x$ in a neighborhood of $\x_o$.
 The multifractal spectrum $D(h)$ is defined as the fractal dimension of the set $\{\x_o | h(\x_o)=h\}$. It can be defined for a measure or for a function.

In practice, estimation of $D(h)$ from its definition is numerically unstable. An alternative approach is to consider 
 the scaling behavior of  the {\em structure functions} $S(q,a)$. More precisely, we look at the multiscaling exponents $\zeta(q)$ such that $S(q,a)\sim a^{\zeta(q)}$, see e.g.~\cite{2004AnGeo..22.2431Y}.  The multifractal formalism is established when one can associate the multifractal spectrum $D(h)$ to its Legendre transform $\zeta(q)$ (in dimension 2):
 \begin{equation}
 \label{Legendretransform}
 	\zeta(q)=2+\inf_{h}[qh-D(h)] \Leftrightarrow D(h) = 2+\inf_q[qh-\zeta(q)]
\end{equation}
 The exponents $\zeta(q)$ reflect the multiresolution statistics of a process~\citep{cgh90,cd95}. It  is used in Section~\ref{S:synthe} as a set of parameters for modeling purpose.

\section{Synthesis of multifractal process}
\label{A:multifrac}

\subsection{Modeling of the images}
A possible approach to model QS EUV images relies on the use of an integrated version of the density of some underlying multifractal measure described by a set of multifractal exponents $\tau(q)$ such that $\tau(0)=\tau(1)=0$. 
 Consider a low-pass filter which multiplies the Fourier transform of the quantity of interest by a factor 
 $1/\|\k\|^{H}$, where $\k$ are the Fourier frequencies. This
 filter modifies the multifractal exponents so that given $\tau(q)$ one gets a measure with multifractal exponents
\begin{equation}
\label{zetaq2}
	\zeta(q)=qH+\tau(q).
\end{equation}
 This is connected to the fact that the differentiation of a H\"older singularity of exponent $h$ becomes a singularity of exponent $h-1$ while its integration yields a singularity of exponent $h+1$~\citep{mh92}.  It is then natural to consider the modeling of Quiet Sun images as a $1/\|\k\|^{H}$ filtered version of a multifractal density. An interesting remark is that $\tau(1)=0$ by definition so that $\zeta(1)=H$. This latter property will be useful to identify the parameters of the proposed models. 
 
\subsection{Compound Poisson Cascades}

Compound Poisson cascades were originally introduced~\citep{baman02} as a multifractal product of cylindrical pulses. This definition can be reformulated as a multiplicative cascade~\citep{cra03a} and extended to dimension $D\geq2$~\citep{c06}. The density $Q_\ell(\x)$ resulting from a compound Poisson cascade is defined by 
\begin{equation}
\label{CPCshaper}
    Q_\ell(\x)=\frac{\prod_{i} W_i^{f(\frac{\x-\x_i}{r_i})}}{\E\left[\prod_{i} W_i^{f(\frac{\x-\x_i}{r_i})}  \right]}
\end{equation}
where $\ell>0$ is some small limiting scale (a kind of resolution); the $W_i$ are independent identically distributed (i.i.d.) non negative random variables called \lq multipliers'; $(\x_i,r_i)$ is a Poisson point process in $\R^2\times [\ell,1]$ with density $dm(\x,r)=(4/\pi r^3) d\x dr$; the function $f(\x)= \1_{[-1/2,1/2]}(\x)$ in the basic definition can be replaced by some compact supported non negative function ( $\1_A$ denotes the indicator function over the set $A$). The integration kernel $f$ plays the role of some geometrical object in the image. It may also be used to attenuate small scales discontinuities or to take into account some geometrical features of the images under study. 

In the limit $\ell\to 0$, compound Poisson cascades are the density of a scale invariant multifractal measure characterized by a set of multiscaling exponents $\tau(q)=q(\E W_i-1)+1-\E W_i^q$ ($\E$ denotes expectation), at least within a certain range of $q$, see equations~\refeq{E:taucpc}-\refeq{ztheoCPC} in Section~\ref{S:identif}. Thus, the design of some $\tau(q)$ function for modeling purpose reduces to the choice of the distribution of the multipliers $W_i$.

An interesting property of the process $Q_\ell(\x)$ is that it can be
interpreted as the intensity $I(\x)$ resulting from the scattering of a uniform light by a random superposition of transparent cylinders of sizes $\{r_i\}$ placed above positions $\{\x_i\}$ and with i.i.d. random transparency $W_i$. The centers $\x_i$ of the cylinders are uniformly distributed on the plane; the radii $r_i$ are distributed by a scale invariant $1/r^3$ law; the distribution of the transparencies $W_i$ is determined by the choice of the function $\tau(q)$ which is directly associated to their second generating function. The intensity of one pixel is therefore the product of the transparencies of the cylinders: this is a multiplicative cascade.  This presentation points to the resemblance between CPC and other classical approaches in image modeling where elementary objects of random sizes are distributed in space following a Poisson point process
\citep{slsz03}. See~\citet{c06,c07pami} for more details.

\section*{Acknowledgements}
{\small The authors would like to thank T. Dudok de Wit and the anonymous referee for their reviewing,  
as well as B. Nicula, O. Podladchikova, and A. Zhukov  for insightful discussions.
Funding of V.D.~and J.-F.H.~by the Belgian Federal Science Policy Office (BELSPO) through the ESA/PRODEX program is hereby appreciatively acknowledged. 
V.D.~thanks the University Blaise Pascal of Clermont-Ferrand for the 1 month stay (may 2005) during which this work was initiated.  This work has been supported by a France-Belgium {\em Hubert Curien} grant ``Tournesol (Communaut\'e francophone)''.}




\begin{table}[htbp]
  \caption{Evolution of the SNR when the distance $d$ to the objective changes, and when the angular resolution $\theta$ increases by a factor $N$, for a uniform source, and a  dirac-like source.
   \label{T:snr_evol}}
  \begin{center}
    \begin{tabular}{|c|| c | c|} \hline
                    SNR      & uniform source & dirac source \\
     \hline \hline
     $d$  changes, &  constant & $\propto$  1/d  \\
     $\theta$ constant & &\\
     \hline
     $\theta$ changes, &   $\propto 1/N $ & constant  \\
     $d$ constant   & &\\
     \hline
    \end{tabular}
  \end{center}
\end{table}

\end{document}